\documentclass[preprint,12pt]{elsarticle}

\biboptions{comma} % Multiple refs will be separated with semicolon

\journal{Physical Review C}

\usepackage{xspace}

\newcommand{\fig}[1]     {Fig.~\ref{#1}}
\newcommand{\figfirst}[1]  {Figure~\ref{#1}}
\newcommand{\tab}[1]     {Table~\ref{#1}}

\newcommand{\secref}[1]  {Sec.~\ref{#1}}

\newcommand{\refref}[1]  {Ref.~\cite{#1}}

\newcommand{\missref}[1]      {\ifshowcomments\textcolor{red}{[REF: #1]}\xspace\fi}
\newcommand{\todo}[1]         {\ifshowcomments\textcolor{red}{[Todo: #1]}\xspace\fi}

\newcommand{\MPPost}          {\texttt{MPPost}\xspace}
\newcommand{\mcnpxpolimi}     {\texttt{MCNPX-PoliMi}\xspace}
\newcommand{\cgmf}            {\texttt{CGMF}\xspace}
\newcommand{\freya}           {\texttt{FREYA}\xspace}
\newcommand{\polimi}          {\texttt{POLIMI}\xspace}
\newcommand{\polimiipol}      {\texttt{MCNPX-POLIMI IPOL(1)=1}\xspace}
\newcommand{\coloronline}     {(Color online)\xspace}

\newcommand{\Dt}{\ensuremath{\Delta t}\xspace}
\newcommand{\Dti}{\ensuremath{\Delta t_i}\xspace}
\newcommand{\Dto}{\ensuremath{\Delta t_1}\xspace}
\newcommand{\Dtt}{\ensuremath{\Delta t_2}\xspace}

\newcommand{\Dtot}{\ensuremath{\Delta t_{1\rightarrow2}}\xspace}

\newcommand{\Eo}{\ensuremath{E_1}\xspace}
\newcommand{\Et}{\ensuremath{E_2}\xspace}
\newcommand{\Emin}{\ensuremath{E_{\textrm{min}}}\xspace}
\newcommand{\Ei}{\ensuremath{E_i}\xspace}

\newcommand{\Eave}{\ensuremath{\overline{E_n}}\xspace}
\newcommand{\Ejave}{\ensuremath{\overline{E_j}}\xspace}

\newcommand{\Si}            {\ensuremath{S_{i}}\xspace}
\newcommand{\Sj}            {\ensuremath{S_{j}}\xspace}
\newcommand{\Dij}            {\ensuremath{D_{ij}}\xspace}
\newcommand{\W}              {\ensuremath{W}\xspace}
\newcommand{\Wij}            {\ensuremath{W_{ij}}\xspace}
\newcommand{\Wtheta}         {\ensuremath{\overline{W}(\theta)}\xspace}
\newcommand{\Wth}[1]         {\ensuremath{\overline{W}(\degrees{#1})}\xspace}
\newcommand{\Asym}           {\ensuremath{A_{\rm{sym}}}\xspace}

\newcommand{\genunit}[2]{\ensuremath{#1~\text{#2}}\xspace}
\newcommand{\meters}[1] {\genunit{#1}{m}}
\newcommand{\cm}[1]     {\genunit{#1}{cm}}
\newcommand{\MeV}[1]    {\genunit{#1}{MeV}}
\newcommand{\ns}[1]{\genunit{#1}{ns}}

\newcommand{\degrees}[1]{\ensuremath{#1^{\mathrm{o}}}\xspace}
\newcommand{\Cftft}{\ensuremath{^{252}\text{Cf}}\xspace}
\newcommand{\Cftfo}{\ensuremath{^{250}\text{Cf}}\xspace}
\newcommand{\Cmtfe}{\ensuremath{^{248}\text{Cm}}\xspace}
\newcommand{\Putf}{\ensuremath{^{240}\text{Pu}}\xspace}

\usepackage[margin=.80in]{geometry}
\usepackage[doublespacing]{setspace}
\usepackage{xcolor}                                     % Extra font colors
\usepackage{amssymb}                                    % Math symbols
\usepackage{amsmath,amsthm}
\usepackage[caption=false,font=footnotesize,subrefformat=parens,labelformat=parens]{subfig}    % Use subfigures
\usepackage{graphicx}                           % The graphicx package provides the includegraphics command.
\usepackage[hidelinks]{hyperref}                        % Links in text
\hypersetup{                    
	colorlinks,
	citecolor=black,
	filecolor=black,
	linkcolor=black,
	urlcolor=black
}
\usepackage{listings}                                   % Box around code
\usepackage{lineno}										  % Add line numbers. 
\newif\ifshowcomments
\showcommentsfalse

\begin{document}
%\tableofcontents

%\SetWatermarkText{DRAFT}
%\SetWatermarkScale{5}
%\SetWatermarkLightness{.93}

%%%%%%%%%%%%%%%%%%%%%%%%%%%%% FRONT MATTER %%%%%%%%%%%%%%%%%%%%%%%%%%%%%%%
\begin{frontmatter}


\title{High resolution measurement of tagged two-neutron energy and angle correlations in \Cftft(sf)}



%%%%%%%%%%%%%%%%%%%%%%%%%%%% AUTHOR INFO %%%%%%%%%%%%%%%%%%%%%%%%%%%%%%

\author[umich]{P.~F.~Schuster\corref{cor1}}
\ead{pfschus@umich.edu}
\author[umich]{M.~J.~Marcath}
\author[umich]{S.~Marin}
\author[umich]{S.~D.~Clarke}
\author[lanl]{M.~Devlin}
\author[lanl]{R.~C.~Haight}
\author[llnl,ucdavis]{R.~Vogt}
\author[lanl]{P.~Talou}
\author[lanl]{I.~Stetcu}
\author[lanl]{T.~Kawano}
\author[lbnl]{J.~Randrup}
\author[umich]{S.~A.~Pozzi}

\cortext[cor1]{Corresponding author}

\address[umich]{Department of Nuclear Engineering and Radiological Sciences, University of Michigan, Ann Arbor, MI 48109, USA}
\address[lanl]{T-2, Los Alamos National Laboratory, Los Alamos, New Mexico, 87545, USA}
\address[llnl]{Nuclear and Chemical Sciences Division, Lawrence Livermore National Laboratory, Livermore, CA 94551, USA}
\address[ucdavis]{Department of Physics, University ofCalifornia, Davis, California 95616, USA}
\address[lbnl]{Nuclear Science Division, Lawrence Berkeley National Laboratory, Berkeley, CA, USA}

%%%%%%%%%%%%%%%%%%%%%%%%%%%%%%%% ABSTRACT %%%%%%%%%%%%%%%%%%%%%%%%%%%%%

\begin{abstract}

\textbf{Background:} Spontaneous fission events emit prompt neutrons correlated with one another in emission angle and energy. \textbf{Purpose:} We explore the relationship in energy and angle between correlated prompt neutrons emitted from \Cftft spontaneous fission. \textbf{Methods:} Measurements with the Chi-Nu array provide experimental data for coincident neutrons tagged with a fission chamber signal with \degrees{10} angular resolution and \ns{1} timing resolution for time-of-flight energy calculations. The experimental results are compared to simulations produced by the fission event generators \cgmf, \freya, and \polimiipol.
\textbf{Results:} We find that the measurements and the simulations all exhibit anisotropic neutron emission, though differences exist between fission event generators.
\textbf{Conclusions:} This work shows that the dependence of detected neutron energy on the energy of a neutron detected in coincidence, although weak, is non-negligible, indicating that there may be correlations in energy between two neutrons emitted in the same fission event.

\end{abstract}

\begin{keyword}
Fission \sep Neutron \sep Correlation

\end{keyword}

\end{frontmatter}


% \include{introduction}
\section{Introduction}

%%% What is the effect we are looking at?
In a fission event, prompt neutron emission occurs on a time scale shorter than that of gamma ray emission~\cite{Fraser1952a,Skarsvag1970}. The emitted neutrons are correlated with one another in their emission angle and energy~\cite{Bowman1962,Larsen2014}. 
%While these correlations have been measured to some degree\missref{Who else has measured this?}, \note{What do we still need to figure out?} This work aims to extend previous measurements in \note{what way?}.
%%% What is the status / what is missing? What could help the situation?
The commonly used \mcnpxpolimi Monte Carlo code treats such correlations using data-based evaluations~\cite{Pozzi2012}. The new, physics-based fission models \cgmf~\cite{Talou2013,Talou2014,Talou2016} and \freya~\cite{Randrup2009,Vogt2011,Vogt2013,Vogt2014,Randrup2014,Vogt2017,Verbeke2018} generate complete events and can thus produce correlations between emitted particles on an event-by-event basis. These codes require high fidelity experimental data for validating their models. 
In this paper, we describe our \Cftft spontaneous fission data, correlated in neutron energy and two-neutron angular separation, and compare the measured correlations to those simulated with fission models \mcnpxpolimi, \cgmf, and \freya, each using \mcnpxpolimi for radiation transport and \MPPost~\cite{Miller2012} for detector response.
% \note{What specific data needs to be improved?}

%%% What are the broader implications for this?
Numerous detector systems exist or are in development for nuclear nonproliferation, safeguards, and arms control applications that would benefit from a better understanding of the correlations in prompt fission neutron emission. One such example is the fast neutron multiplicity counter, a nuclear safeguards instrument that is used for nondestructive assay of special nuclear material~\cite{Frame2007,DiFulvio2017}. 
% Compared to traditional neutron multiplicity counters that measure thermal neutrons, fast neutron multiplicity counters enable more precise measurements of the neutron interaction time.
% Fast neutron multiplicity methods in general. Provide a few references. Ours on Plutonium INL measurements with FNPC. IAEA may have some references for their systems (fast n), based on liquid scintillator. \note{Just a couple of sentences, not a long discussion. Ask Mike Hamel about IAEA references}
Similarly, applications have been proposed for exploiting the correlations that exist between neutrons emitted from the same fission event in multiplying materials where fission chains are present~\cite{Mueller2016,Holewa2013}. 
%For example, Mueller and Mattingly demonstrated that the neutron emission anisotropy can be used as a means of quantifying the multiplicity of an unknown assembly of plutonium metal without revealing other sensitive information, which may be useful for an arms control application~\cite{Mueller2016}. % This work demonstrated that no anisotropy is observed for sources with multiplicity greater than 3. In an arms control verification measurement, in which one wishes to verify the presence of highly multiplying material without revealing other sensitive information about the material, this technique could serve as a physics-based zero knowledge measurement. Another application was explored by 
%Additionally, Holewa \etal demonstrated that the neutron angular anisotropy can be used to dynamically determine $\alpha$, the ratio of the $(\alpha,n)$ rate to spontaneous fission~\cite{Holewa2013}. This is useful in coincidence counting applications when only singles and doubles information is available (triples unavailable). 
Accurate physics models are important in the development of these systems and methods.

%%% Previous work
\todo{Include previous work. What measurements have been made in the past? How will our measurements differ?}

%%% What are we going to do and how?
This paper presents measurements and simulations of correlated neutrons from \Cftft spontaneous fission to confirm and extend previously reported results. Measurements were made with 42 detectors of the Chi-Nu detector array at the Los Alamos Neutron Science Center (LANSCE) at Los Alamos National Laboratory (LANL)~\cite{Perdue2014}. The Chi-Nu array covers a large solid angle with detectors approximately \meters{1} from the source, thereby providing high efficiency and excellent timing resolution for time-of-flight energy calculations. Additionally, the \Cftft source was embedded in a fission chamber, providing good time resolution for the fission event signal. Double coincident neutron events, in which two neutrons are detected in coincidence with a fission chamber trigger, were identified as ``bicorrelation'' events, as explained in \secref{sec:bicorr_events}. 
The measurement offers improved angle resolution, excellent timing resolution, and enhanced background suppression compared to previous work~\cite{Larsen2014,Verbeke2018b}. A previous paper by the authors investigated correlations between the prompt neutron and photon multiplicities~\cite{Marcath2018}. This work includes the first comparison of correlated neutron energy characteristics for \Cftft spontaneous fission, including a new observable: the average energy of neutrons detected in coincidence with emitted neutrons at a given energy as a function of the angle between them.

% Our measurement system covers more solid angle with larger detectors. Better efficiency. Using PSD capable detectors. We have simulation experience that allows us to model the detector response. We also have a fission chamber. Lots of data. Therefore we have good statistics on higher order events.

%\note{What makes this paper so important is what is novel about these measurements vs. what we have done in the past and vs. previous publications. Need to make it clear to the reviewers-MM}

% \input{experimental}

\section{Experimental, Simulation, and Analysis Methods}

\subsection{Measurement Setup and Methods}

\begin{figure}[h]
	\centering\includegraphics[scale=1]{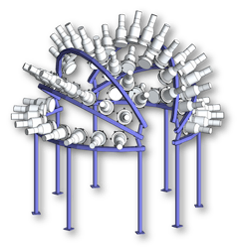}
	\caption{Diagram of the Chi-Nu detector array. \todo{Ask for image with better resolution}}
	\label{fig:chi_nu_diagram}
\end{figure}

\begin{figure}[!t]
	\centering
	\includegraphics[trim={0cm 0cm 0cm 0cm},clip,scale=1]{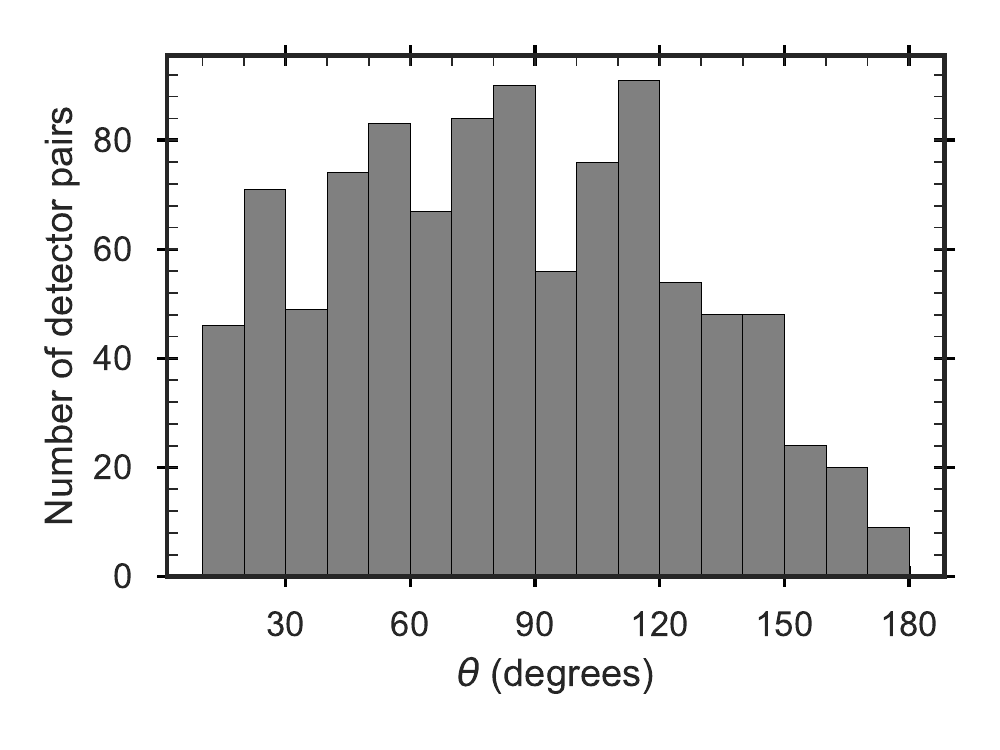}%	
	
	\caption{The number of detector pairs contributing to each \degrees{10} bin is shown.}
	\label{fig:det_df}
\end{figure}

In this work, we employ the data taken with the Chi-Nu detector array, illustrated in \fig{fig:chi_nu_diagram}, at the LANL LANSCE facility in 2015, using a \Cftft spontaneous fission source, for our bicorrelation analysis. Because the experimental setup for this analysis was described in detail in \refref{Marcath2018}, we only briefly summarize the parts of the setup relevant for this analysis here. 
%An illustration of the detector array is shown in~\cref{fig:chi_nu_diagram} and 
The Chi-Nu array consists of 54 EJ309 liquid scintillator detectors mounted at \degrees{15} intervals along six arcs to form a hemispherical distribution of detectors. Each detector is cylindrical, \cm{17.78} in diameter and \cm{5.08} thick, coupled to a \cm{12.7} diameter photomultiplier tube (Hamamatsu R4144). Each detector subtends approximately a \degrees{10} angle from the source. % \note{How is d of det $>$ pmt? Light guide?} 
Limitations in the acquisition system constrained these measurements to using only 42 detectors, making for 861 pairs of detectors at angles from \degrees{15} to \degrees{180}. The large number of detector pairs produces a wide range of angles, allowing for discretization in \degrees{10} bins, as shown in \fig{fig:det_df}, which is improved compared to previous work that only had \degrees{30} or \degrees{90} resolution~\cite{Larsen2014,Mueller2016}. Each \degrees{10} bin contains multiple detector pairs; observable quantities are averaged across all pairs within a bin to reduce statistical error per bin. 

A \Cftft(sf) source was embedded in a fission chamber, with characteristics detailed in~\refref{Marcath2018}. The source was fabricated in 2010 and the measurements reported here were performed in 2015. The \Cftft spontaneous fission count rate was $2.98\times10^5$ spontaneous fissions per second, with negligible contributions from spontaneous fission of \Cftfo and \Cmtfe. The source was placed at the focal point of the hemispherical array so that detectors were approximately \meters{1} from the source. Over the duration of the measurement, $2.2\times10^9$ fission events occurred, resulting in $1.42\times10^9$ fission chamber triggers.
% Note: Include count rate, threshold, measurement time, number of fissions observed, number of doubles observed?

The use of the fission chamber makes this measurement unique compared to similar measurements\missref{which?} made in the past because it provides a reference time for when a fission event occurs. Thus the neutron time-of-flight may be directly calculated for each detected neutron whereas previous work was limited to calculating the difference between the detection times of correlated particles~\cite{Larsen2014}. %Some previous measurements did include a fission chamber \missref{Which ones?}, but they differed because \todo{how?}. 

% \begin{figure}[h]
% \centering\includegraphics[width=0.8\linewidth]{fig/array_photo.jpg}
% \caption{Photo of the Chi-Nu detector array.}
% \label{fig:chi_nu_photo}
% \end{figure}

Full waveforms were recorded with three CAEN V1730 digitizers with 500 MHz sampling and 14-bit amplitude resolution over a 2 V range and post-processed in digital form. Standard digital pulse processing was implemented, as detailed in~\refref{Marcath2018}. Particle types were classified using charge-integration $n$-$\gamma$ pulse shape discrimination (PSD)~\cite{Polack2015}, which was performed offline and optimized uniquely for each detector. 
A quadratic PSD line was used to discriminate between the neutrons and photons with misclassification of low light output events estimated to be approximately 1\% of all measured events. 

The measurement had a pulse height threshold of 100 keVee (``electron-equivalent'' keV) light output, corresponding to approximately \MeV{0.8} neutron energy deposited. This threshold was selected to minimize misclassification of photons as neutrons in the measurement, which mostly occurs below this threshold. An upper voltage limit reduced the experimental sensitivity to neutrons with energy depositions above \MeV{8.1}. This work focuses on events where both detected neutrons have energies in the range \MeV{1} to \MeV{4}, due to reduced statistics at higher energies.

%In the paper: Include a description of PSD misclassification, how it varies with $Delta_tn$ and therefor $E_n$
%Put some boundaries on misclassification
%Compare to other data from the literature
%Be clear that threshold is from TOF, better than pulse height threshold

% \input{simulation}
\subsection{Simulation Techniques}

The measurement set-up was simulated using \mcnpxpolimi, which models the laboratory geometry and performs the particle transport. The system was modeled in great detail, including the Chi-Nu structure, concrete floor, and fission chamber. Waveform processing and particle-type classification is assumed to be perfect in the simulation so that all events are identified as the correct particle type. A light output threshold of 100 keVee was used to match that of the experimental data. 

In order to study different fission models, \cgmf, \freya, and the built-in PoliMi source \texttt{IPOL(1)=1}, referred to as \polimi, were used in \mcnpxpolimi. These models produced list-mode data including initial energy, initial direction, and particle type for each particle generated in an individual fission event, which was passed to \mcnpxpolimi for transport. Following transport, \mcnpxpolimi produced a file with event-by-event information on interactions in detector cells. Detector response was calculated with \MPPost post-processing software~\cite{Miller2012}, which handles the nonlinear light output of organic scintillators.

% Geant paper 2012 JNMM

\polimi and \freya simulated $10^9$ fission events, while $1.92\times10^8$ \cgmf events were employed, resampled with new, randomly sampled, fission fragment directions from a subset of $1.92\times10^6$ events. \tab{tab:counts} shows these values and the number of detected bicorrelation events in all four datasets.

\begin{table}[ht]
	\centering
	\caption{Experimentally detected and generated fission events resulting in the given total and per fission bicorrelation counts.}
	\label{tab:counts}
	\begin{tabular}{l|c|c|c}
		&number fissions& bicorrelation counts & bicorrelation counts per fission \\
		\hline
		Experiment   & $1.42\times10^9$ & $(3.941\pm0.002)\times10^6$ & $(2.771\pm0.002) \times10^{-3}$ \\
		\cgmf         & $1.92\times10^8$ & $(0.737\pm0.085)\times10^6$ & $(1.786\pm0.004) \times10^{-3}$\\
		\freya        & $1.00\times10^9$ & $(2.978\pm0.002)\times10^6$ & $(2.978\pm0.002) \times10^{-3}$\\
		\polimi       & $1.00\times10^9$ & $(3.409\pm0.002)\times10^6$ & $(3.409\pm0.002) \times10^{-3}$       
	\end{tabular}
\end{table}

\subsection{Identifying Bicorrelation Events}\label{sec:bicorr_events}

\begin{figure}[h]
	\centering\includegraphics[width=0.5\linewidth]{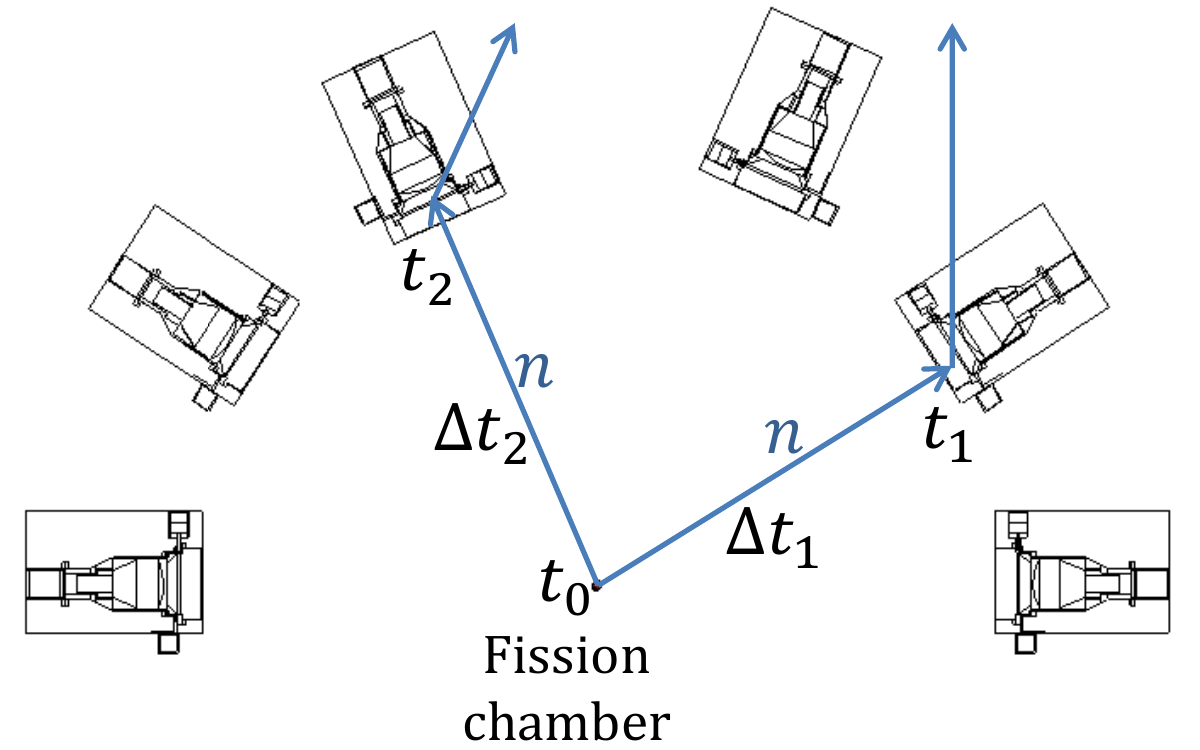}
	\caption{\coloronline Schematic of a true bicorrelation event in which two prompt fission neutrons are detected in coincidence with their originating fission. The schematic used is a two-dimensional view through an arc of the detector array in the \mcnpxpolimi model. }
	\label{fig:sketch}
\end{figure}

This paper studies the relationship between pairs of detected neutrons that are emitted from the same fission event, as illustrated in~\fig{fig:sketch}. The interaction times of two neutrons, $t_1$ and $t_2$, were each correlated with the corresponding fission chamber trigger time, $t_0$, in the measured data, and the time of flight of each neutron was calculated as $\Delta t_i = t_i-t_0$. In the \mcnpxpolimi simulations, the times of flight were provided directly on an event-by-event basis in the output file. These double neutron events are referred to as ``bicorrelation'' events, a term first applied to coincident radiation counting by Mattingly~\cite{Mattingly1998}, because both detected neutrons are correlated with the fission chamber trigger. Bicorrelation events were selected as any double neutron interaction within \ns{200} of the fission time. If $n>2$ prompt neutrons were detected, $n\choose2$~separate bicorrelation events were recorded: one from each pair of detectors. For example, if three neutrons were detected from the same fission event, then three pairs of neutrons were analyzed.

%\note{Present or past tense?} 
Bicorrelation events include interactions of prompt fission neutrons that travel straight to the detector, which are the \textit{true} bicorrelation events, and events in which one or both of the neutrons comes from an accidental interaction such as room return, cross talk, and, in the case of experimental data, natural background. In our experiment, triggering in coincidence with the fission chamber offers significant background suppression compared to other measurements that do not use a fission chamber signal. Background in bicorrelation events in this measurement was estimated to be less than 2.5\% of the overall signal and had a negligible effect on the final results. %While it is possible to subtract background in the time-of-flight distributions explained in the next section, it is not possible to subtract background in the energy distributions, and therefore background subtraction was not performed in the analyses presented here.

% However, this analysis uses several techniques to minimize contributions from these accidental effects. Thus, any such event in which an interaction is detected in the fission chamber along with a double neutron detection is referred to as a bicorrelation event. 

This work will study the characteristics of bicorrelation events with respect to the angle between the neutrons; this will be referred to as the bicorrelation angle and is approximated as the angle between the centers of each detector with respect to the fission chamber.

% These two or more interactions in the detectors are correlated in time to the event in the fission chamber, and therefore are likely to be particles emitted by the fission event. However, other accidental events including background may be observed. Room return is also possible, in which a fission-emitted neutron is observed after it has scattered off of materials in the room. Cross-talk is also observed, which is the coincident measurement of a single neutron (likely a fission neutron) that interacts in two detectors, thereby appearing like a pair of neutrons. 

\subsection{The Bicorrelation Distribution}\label{sec:bicorrelation_distribution}

\begin{figure}[!t]
	\centering
	\subfloat{\includegraphics[trim={0cm 0.5cm .5cm .5cm},clip,width=3in]{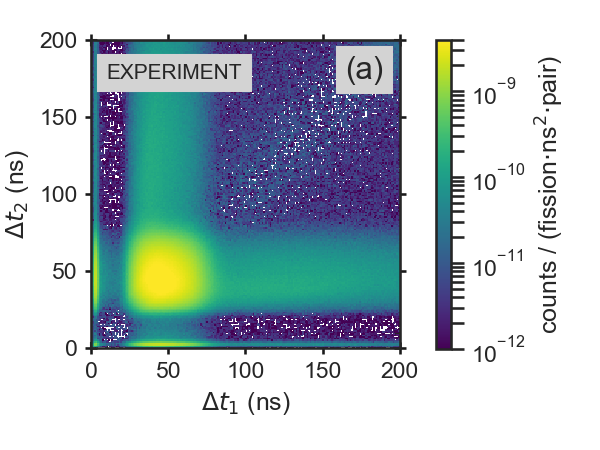}}
	\hfil
	\subfloat{\includegraphics[trim={0cm 0.5cm .5cm .5cm},clip,width=3in]{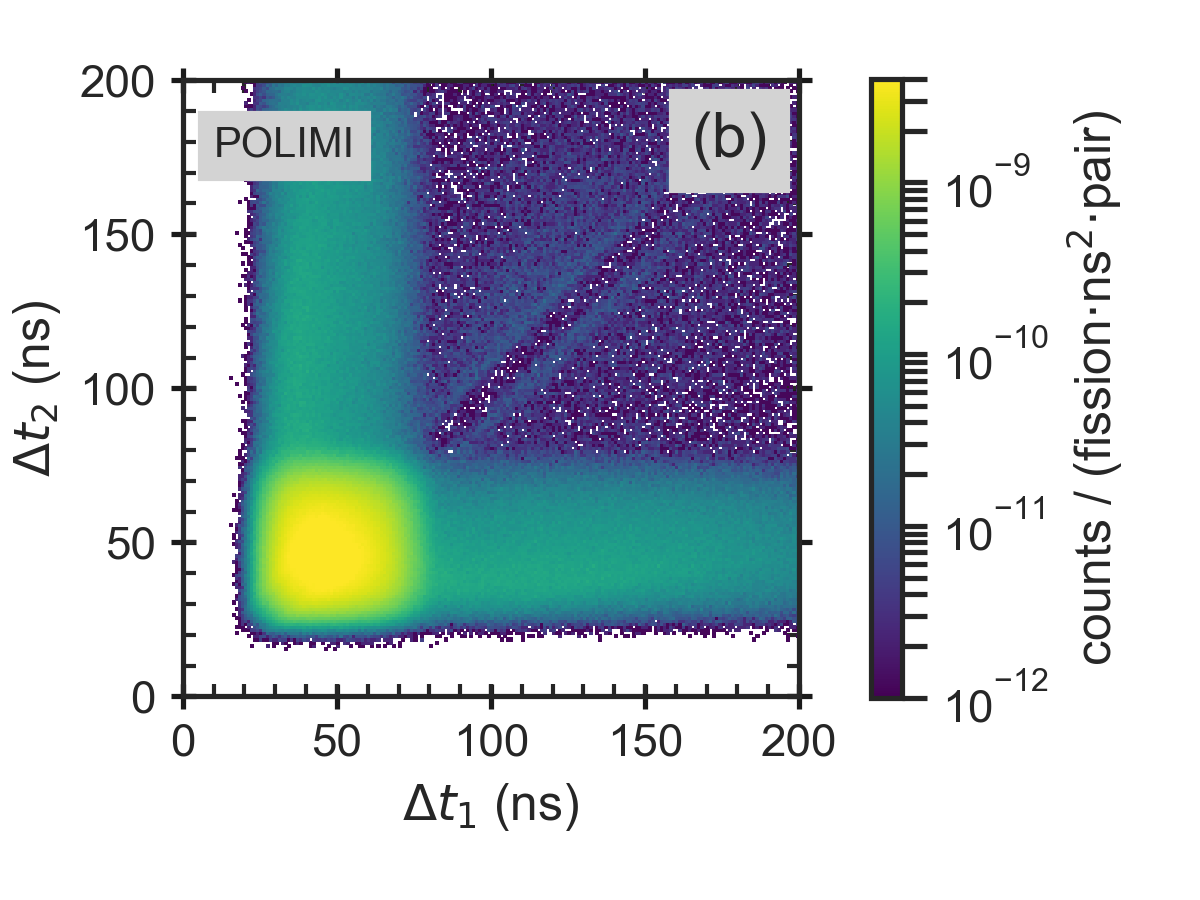}}
	\hfil
	\subfloat{\includegraphics[trim={0cm 0.5cm .5cm .3cm},clip,width=3in]{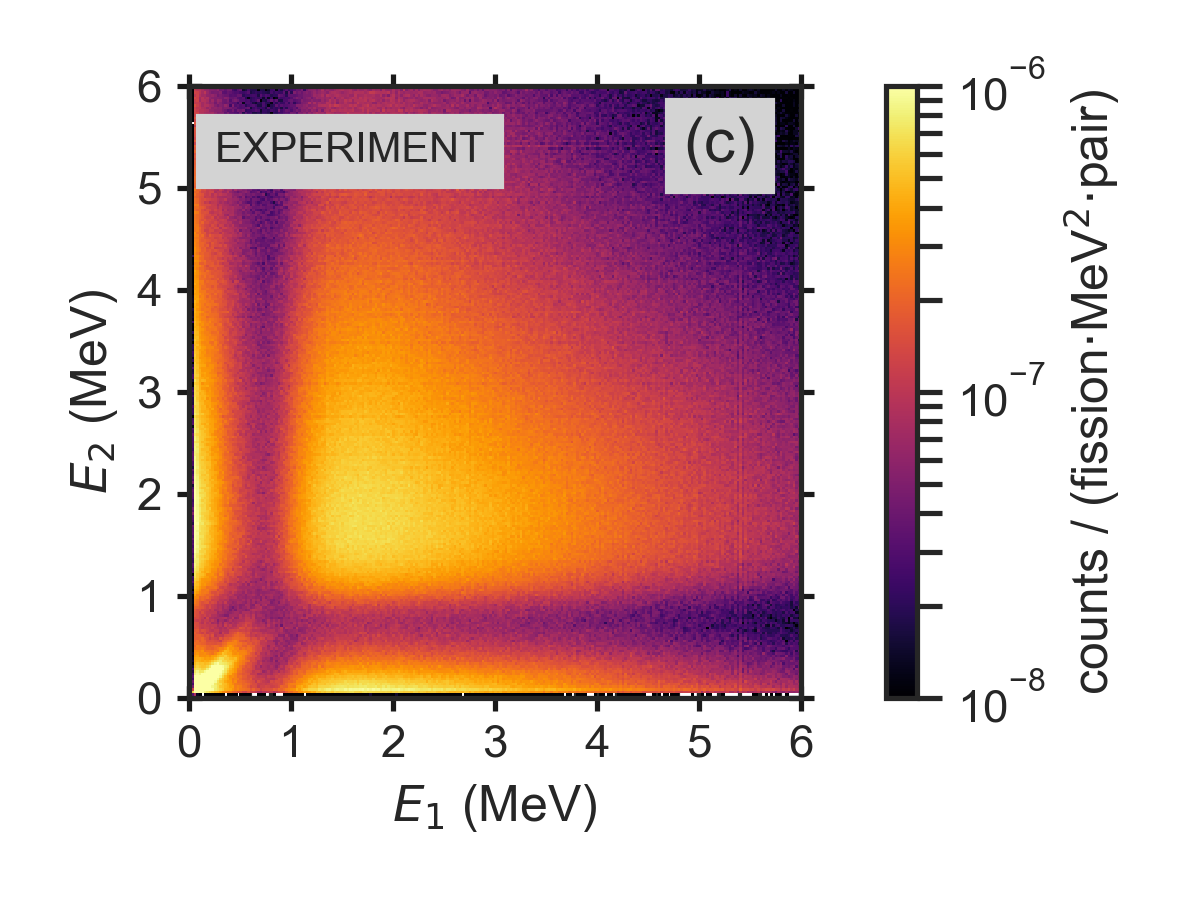}}
	\hfil
	\subfloat{\includegraphics[trim={0cm 0.5cm .5cm .3cm},clip,width=3in]{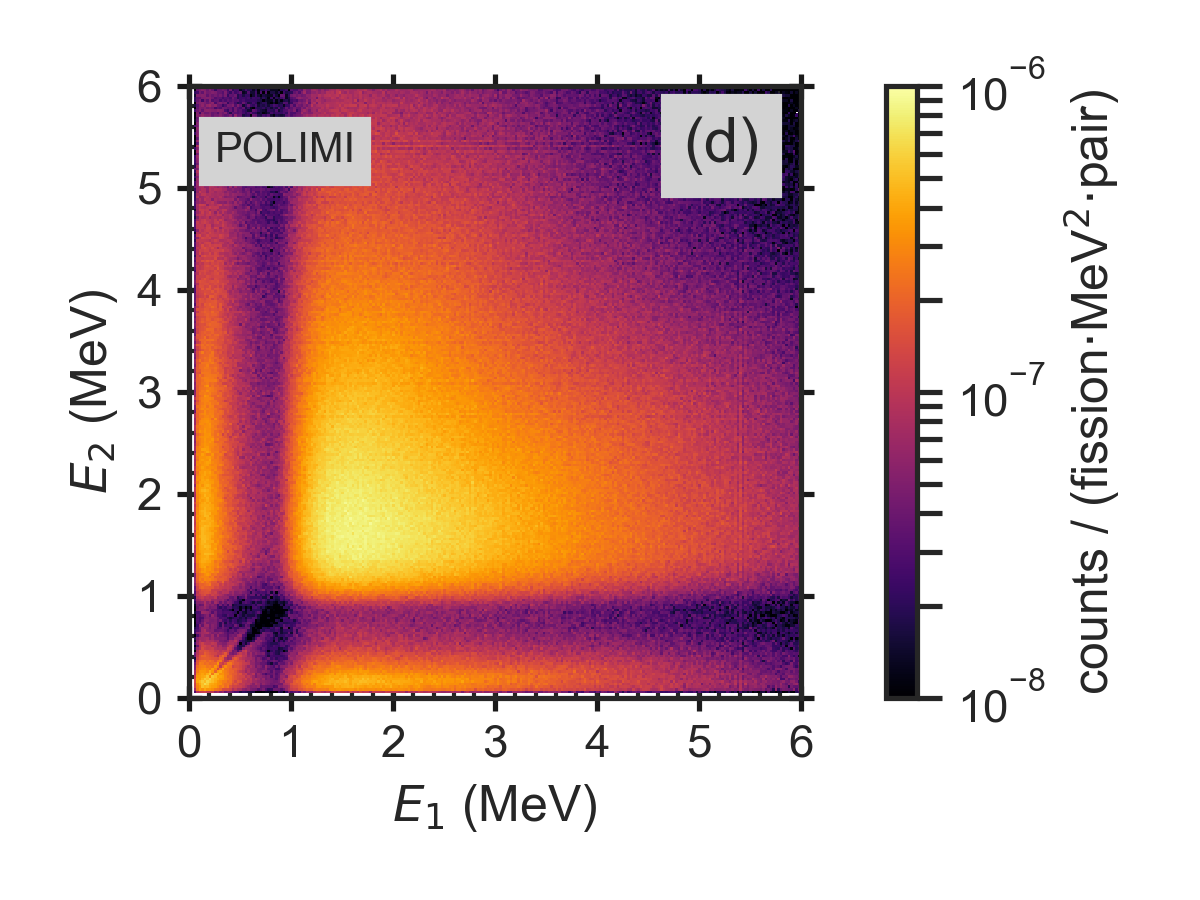}}
	\caption{\coloronline Bicorrelation time-of-flight distribution for (a) experimental data and (b) \polimi simulations, and bicorrelation energy distribution for (c) experimental data and (d) \polimi simulations.}
	\label{fig:bicorr}
\end{figure}

This analysis will make use of the bicorrelation distribution: a two-dimensional distribution of time of flight or energy for bicorrelation neutron events. The energies are calculated from the times of flight with the assumption that the neutron traveled directly from the fission chamber to the detector. Slight differences in the distances from the fission chamber to each detector are incorporated. \figfirst{fig:bicorr} shows the bicorrelation distributions for the experiment and \polimi simulations. These distributions show the number of counts at each $(\Dto,\Dtt)$ or $(\Eo,\Et)$ pixel, normalized by the number of detector pairs, the number of fission events, and the pixel size. 

There are many interesting features in these distributions, a few of which are described here. The first observation is the primary feature produced by true prompt fission neutron bicorrelation events. In the time-of-flight distributions, this feature appears as a bright yellow spot within the approximate time window $25 < \Dti < \ns{75}$. In the energy distributions, this feature appears as a bright distribution extending to larger neutron energies from approximately $\Ei=\MeV{1}$ for each neutron, which corresponds to the peak of the prompt fission neutron spectrum.

A second feature that can be observed in the bicorrelation distributions is the presence of accidental events, such as room return. In the time-of-flight distributions, these events appear at times beyond the true bicorrelation region, and dominate at $\Dti > \ns{75}$, where double-accidental events exist. Events in which a single accidental neutron is detected in coincidence with a true prompt fission neutron produce the wide bands emanating from the true bicorrelation region toward higher \Dt. When converted to neutron energy, these long time-of-flight events are mapped to very low energies and appear on the bicorrelation energy distribution as the bright yellow regions along the axes and as a bright spot at the origin. 

A third feature visible in the experiment time-of-flight distributions is PSD misclassification, which appears as the narrow bands along the $x$- and $y$-axes and as a localized spot at the origin. In this case, one or both of the particles is a gamma ray with a very small \Dti that has been misclassified as a neutron. While this feature in the time-of-flight distribution is very similar to the accidental event features in the energy distribution, they are, in fact, different. This feature due to misclassification does not appear in the \polimi distribution, as all simulations assume perfect PSD and thus do not include misclassified events. 

A final feature that is barely visible in these time-of-flight distributions is cross talk. This effect is explored in more detail in the next section.

\subsection{Cross Talk}
Cross talk occurs when the same neutron interacts in multiple detectors and produces a false bicorrelation event. Cross talk is prevalent in detector pairs with small angular separation. Because full simulations were performed for all fission event generators, cross-talk events are present in all simulations and in the experimental data. %  are accidental bicorrelation events, and are not representative of true prompt fission bicorrelation neutrons. 
Although it is possible to remove cross talk on an average basis, as performed in~\refref{Marcath2016}, there is no way to remove cross talk on an event-by-event basis in experimental data\missref{Desesquelles1990}. Therefore, cross-talk events and their effects on the bicorrelation analysis are present and will be discussed throughout this work.

\begin{figure}[!t]
	\centering
	\subfloat{\includegraphics[trim={0cm 0.5cm .5cm .5cm},clip,width=3in]{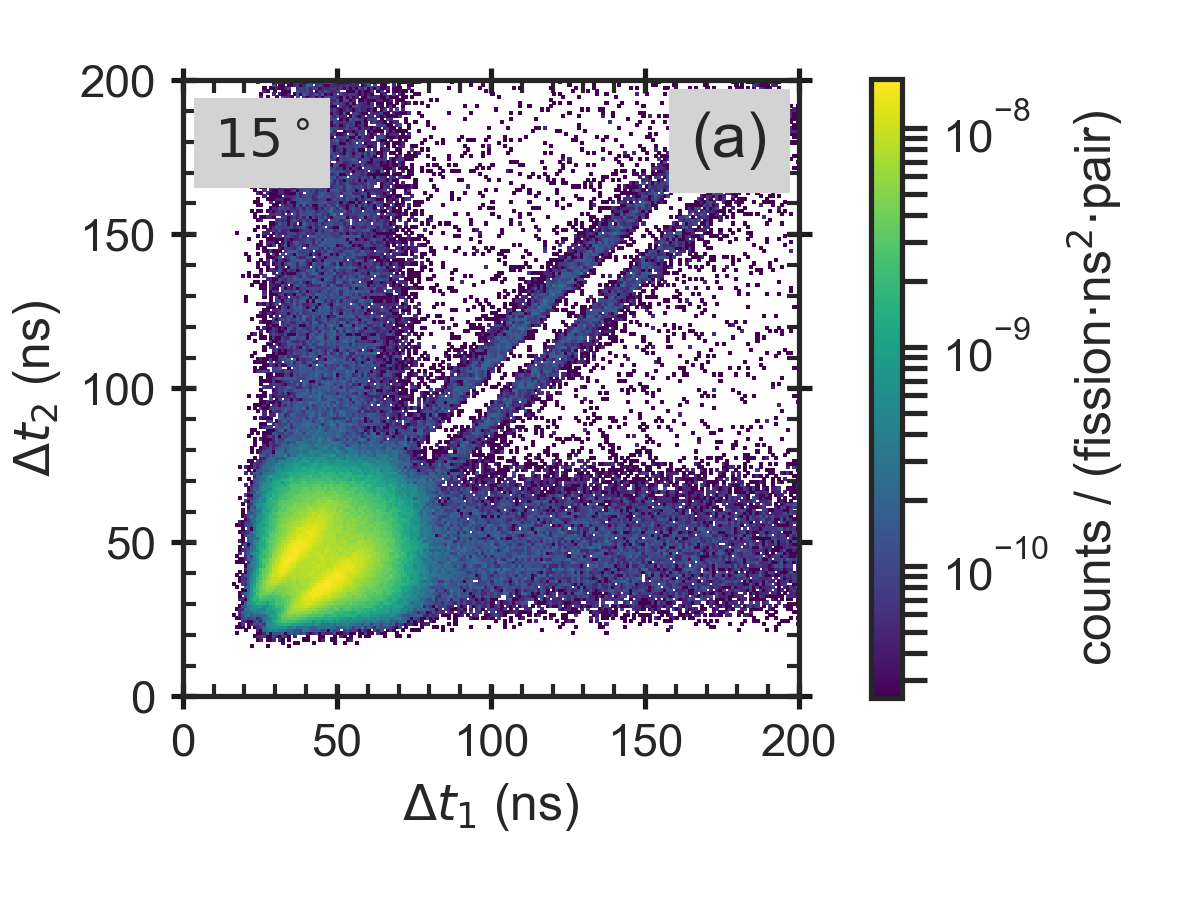}}
	\hfil
	\subfloat{\includegraphics[trim={0cm 0.5cm .5cm .5cm},clip,width=3in]{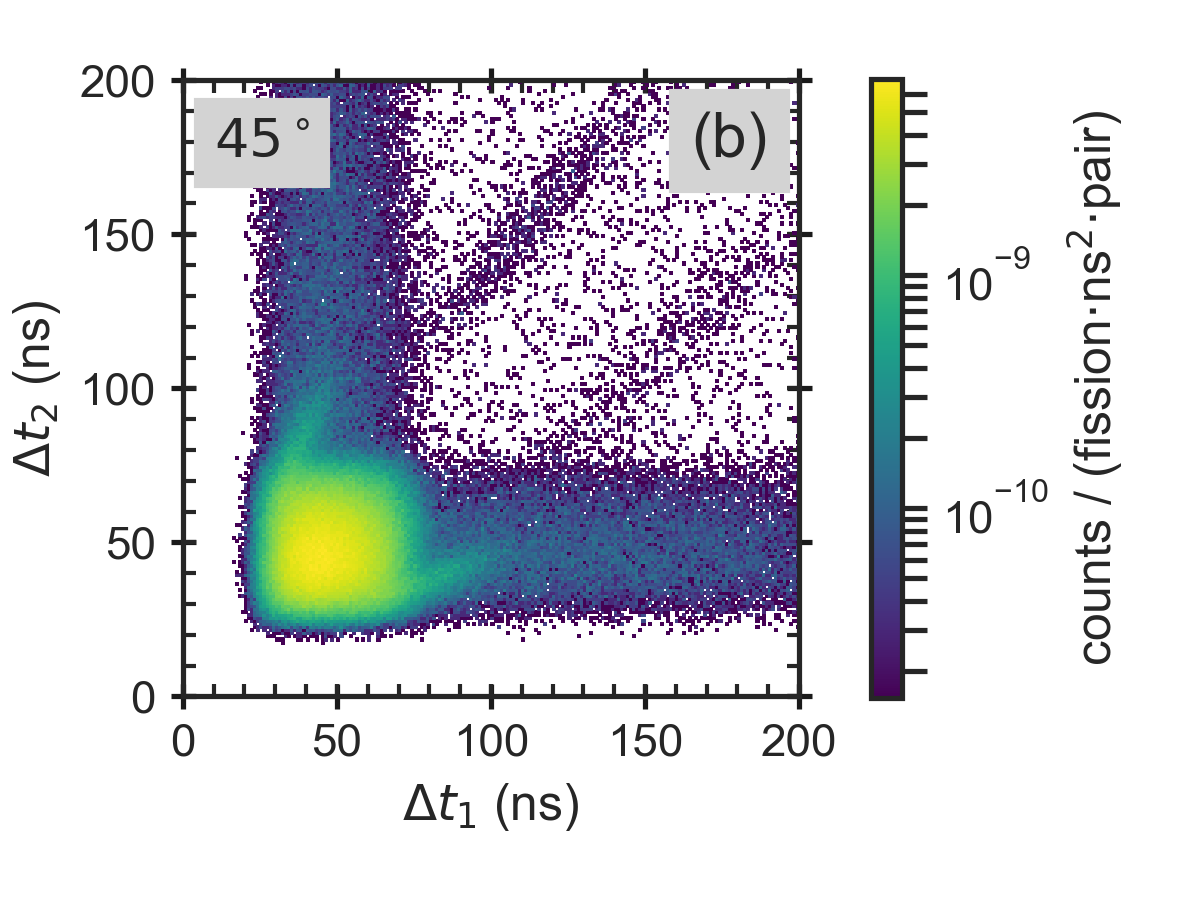}}
	\hfil
	\subfloat{\includegraphics[trim={0cm 0.5cm .5cm .5cm},clip,width=3in]{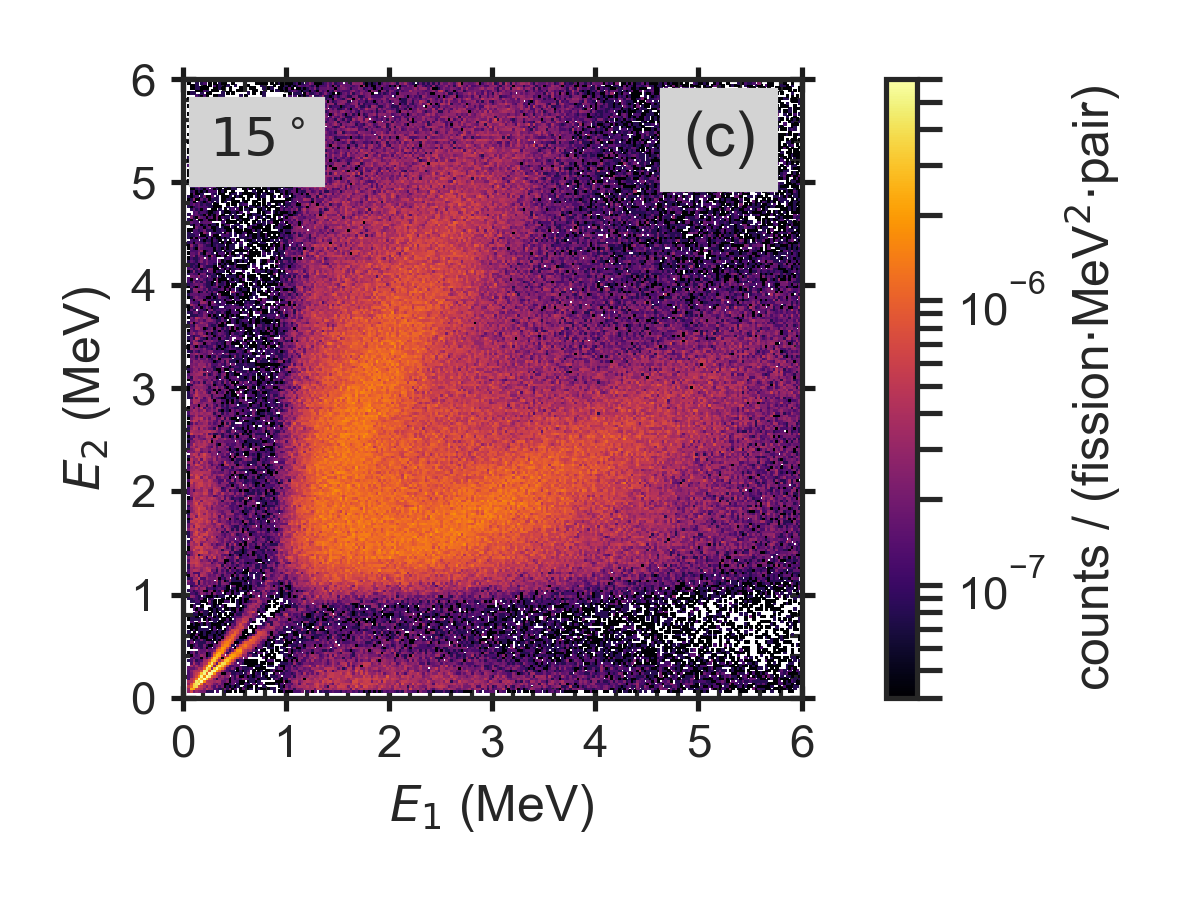}}
	\hfil
	\subfloat{\includegraphics[trim={0cm 0.5cm .5cm .5cm},clip,width=3in]{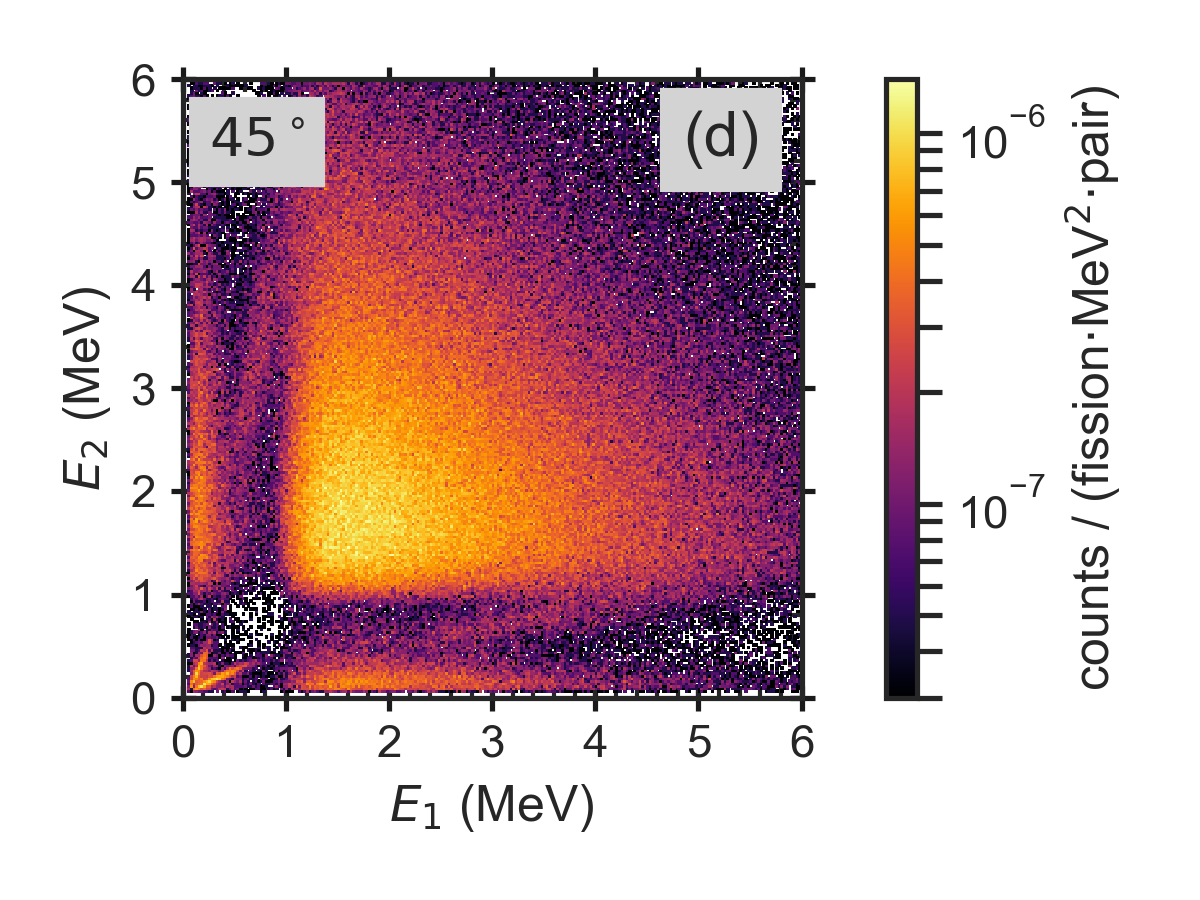}}
	\caption{\coloronline Bicorrelation (a,b) time-of-flight and (c,d) energy distributions from \polimi simulation showing cross-talk effects, displayed for  detector pairs at (a,c) \degrees{15} and (b,d) \degrees{45}. The diagonal bands in each distribution include cross-talk events, which move farther from the identity line ($\Dto=\Dtt$ and $\Eo=\Et$) and decrease in magnitude as the angle between detectors increases. }
	\label{fig:crosstalk}
\end{figure}

Cross-talk events can be visually identified on small-angle bicorrelation distributions, as shown in \fig{fig:crosstalk} for a \polimi simulation of detector pairs at \degrees{15} and \degrees{45}. Cross-talk events appear as two diagonal bands in the bicorrelation time-of-flight and energy distributions. The line of cross talk can be defined as $\Dtt = \Dto + \Dtot$ when the neutron interacts first in detector 1, and \Dtot is the time-of-flight between detectors. Likewise, $\Dto = \Dtt + \Dtot$ describes the line of cross-talk events in which the neutron interacted first in detector 2. Then \Dtot will follow a distribution according to the energies of neutrons traveling from detector 1 to detector 2 and the distance between them. Thus, the $(\Dtt,\Dto)$ distribution will be diagonal lines with widths determined by the \Dtot distribution and offset from the identify line $\Dtt = \Dto$ by the magnitude of \Dtot. As the angle and distance between detectors increases, \Dtot increases and the cross-talk bands decrease in magnitude and departs farther from the identity line $\Dtt = \Dto$. The cross-talk features at long times ($>\ns{75}$) and low energies ($<\MeV{1}$) are largely due to accidental cross-talk events from room return or background. 

Cross-talk effects are prominent at \degrees{15} and visible in some distributions up to \degrees{75}. Regions that may be affected by cross talk are displayed with gray background in the analysis plots in the next section.

\section{Analysis and Results}

\subsection{Anisotropy in Neutron Emission Rate}\label{sec:asym}

Neutron emission from \Cftft spontaneous fission is assumed to occur after the fission fragments are in motion and traveling in opposite directions~\cite{Vorobyev2009}. This does not apply to scission neutrons, which are assumed to be emitted isotropically in the rest frame of \Cftft prior to the full acceleration of the fragments. We note that even if scission neutrons are emitted isotropically, they may not be detected as such in the lab frame. Scission neutrons make up an unknown fraction of the total neutron emission, estimated at approximately $10-20\%$~\cite{Vogt2014,Petrov2005,Chietera2018}.

In our simulations, we assume that neutron emission is isotropic in the rest frame of each fission fragment but anisotropic in the laboratory frame of motion. Thus, the direction of neutron emission follows that of the fission fragment that emitted it so that neutrons emitted in the direction of the fission fragment will receive an energy boost. The anisotropy can be characterized by calculating the count rate of bicorrelation events in detector pairs as a function of bicorrelation angle. The relative bicorrelation count rate, \Wij, for each pair of detectors $i$ and $j$, is defined as~\cite{Mueller2016}:

\begin{equation}
\label{eq:Wij}
\Wij = \frac{\Dij}{\Si \Sj},
\end{equation}
where \Dij is the doubles count rate, and \Si and \Sj are the corresponding singles count rates. Each of these rates are determined from the number of counts in the energy range \MeV{1} to \MeV{4}. This conservative energy range was selected to minimize threshold effects at low energies and gamma-ray misclassification at higher energies while maximizing statistics. This analysis corrects for slight variations in efficiency between detectors.

An average \W was calculated for detector pairs in each \degrees{10}  bin, \Wtheta. The angle is plotted at the midpoint of the bins. For example, the data points at \degrees{15} include all pairs in the range $(\degrees{10},\degrees{20}]$. The error in \Wtheta was calculated as the standard deviation of $W$ values in that angular range. This error is larger than the propagated statistical error and attempts to incorporate systematic errors. 

\begin{figure}[!t]	
	\centering
	\subfloat{\includegraphics[trim={0cm 0cm 0cm 0cm},clip,width=3in]{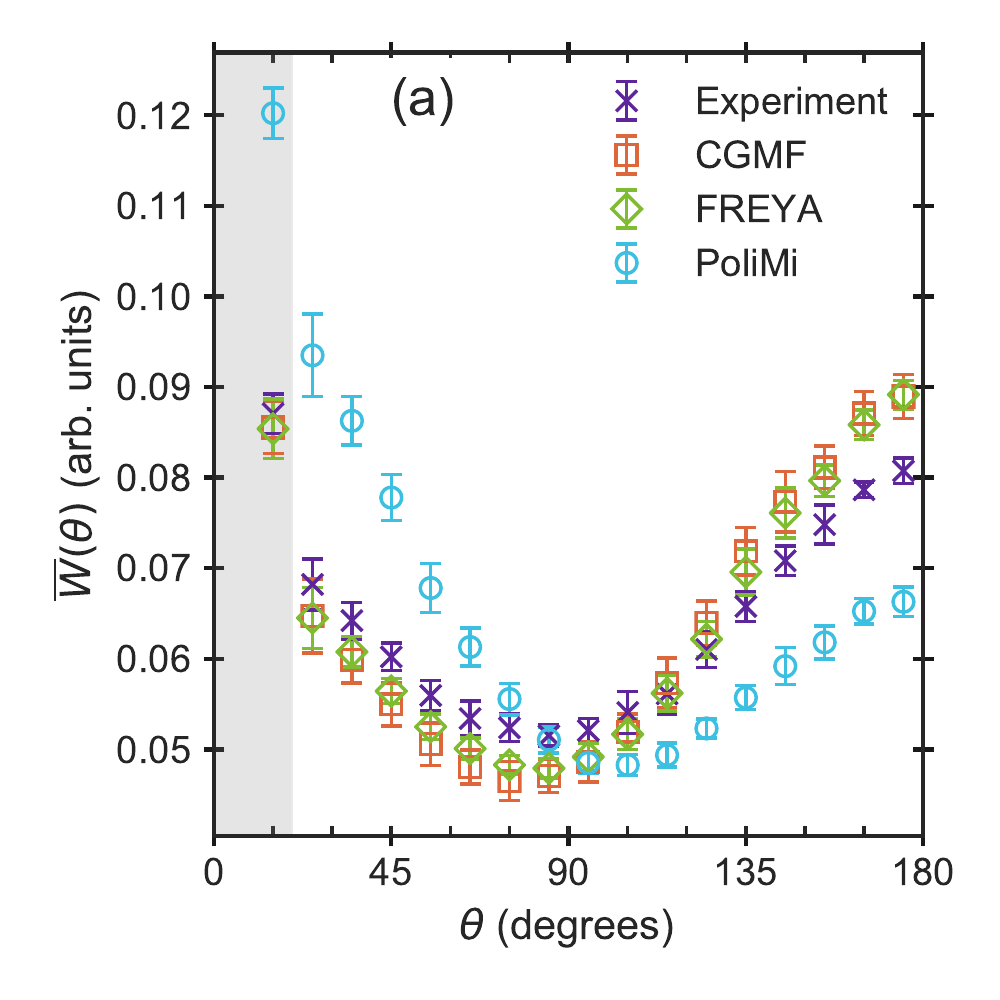}}
	\hfil
	\subfloat{\includegraphics[trim={0cm 0cm 0cm 0cm},clip,width=3in]{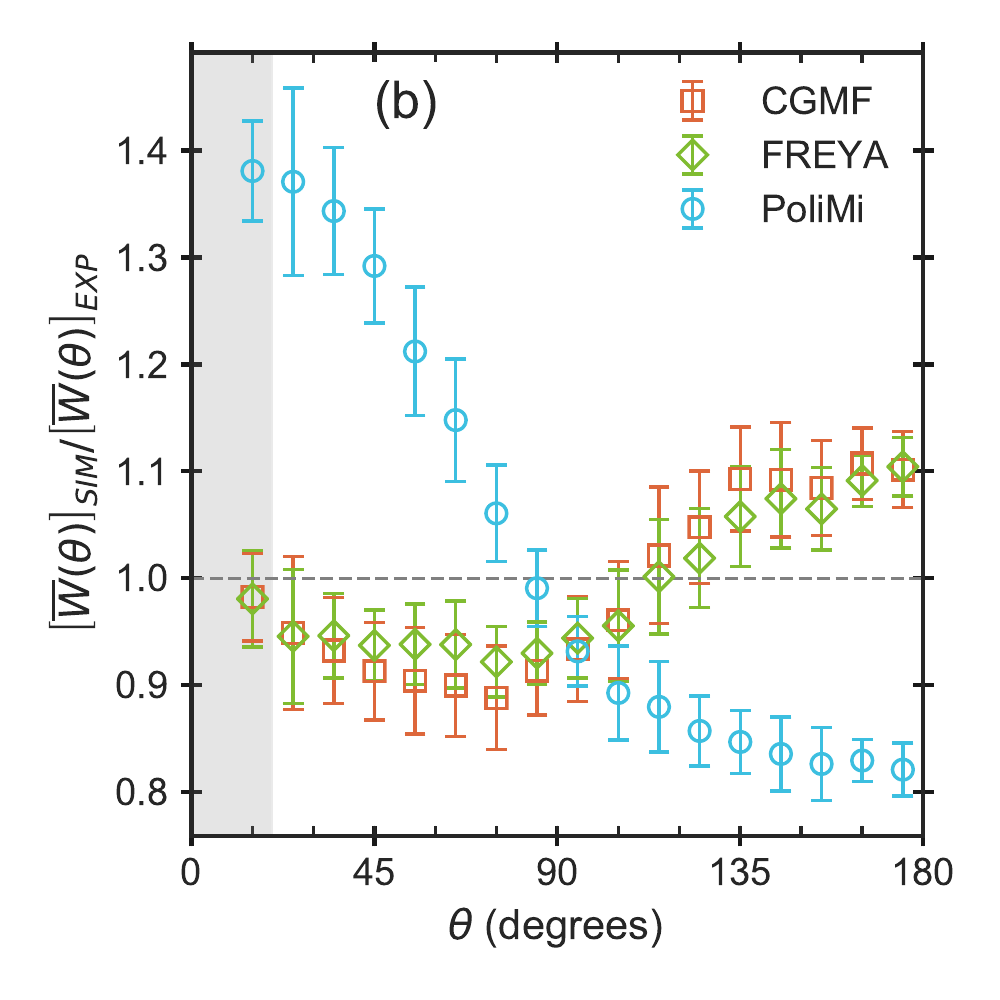}}
	\caption{\coloronline (a) Relative bicorrelation count rate \Wtheta for events from \MeV{1} to \MeV{4}, normalized by integral. (b) Ratio between relative bicorrelation rate for each simulation to that from experiment. The gray region from \degrees{0} to \degrees{20} serves as a reminder that cross talk is significant over this range.}
	\label{fig:W}
\end{figure}

\figfirst{fig:W}(a) shows \Wtheta for all four data sets, normalized by the integral over the distribution. All four datasets produce smoothly varying distributions with a local maximum at \degrees{15} where cross talk is prevalent, a minimum near \degrees{90}, and a local maximum at \degrees{175}. The minimum angle varies from \degrees{75} for \cgmf to \degrees{85} for \freya and the experiment to \degrees{105} in \polimi. The experimental result and the \polimi simulation agree within uncertainties with previous work with lower angular resolution~\cite{Larsen2014}. The largest magnitude of change between $\Wth{175}$ and $\Wth{85}$ is found with \cgmf, while the smallest magnitude of change is seen in \polimi. 

The most striking difference is that the \polimi result is tilted to the left, while the \cgmf, \freya, and experimental results are tilted to the right. The tilt of the angular correlation is strongly tied to the sharing of angular excitation energy between fission fragments. The complete event models \cgmf and \freya handle this sharing and give some additional energy to the light fragment. In \freya, this is done with the $x$ parameter, defined as the advantage in excitation energy given to the light fragment~\cite{Vogt2014}, where $x$ is an adjustable input parameter expected to be larger than 1. The best fit value of $x$ for \freya with \Cftft(sf) was found to be 1.27~\cite{VanDyke2018}. A larger value of $x$ makes the distribution tilt toward \degrees{0} as the light fragment receives more energy and emits more neutrons, increasing the zero degree correlation. A value of $x$ near 1 makes the distribution tilt more strongly toward \degrees{180} as the energy is split more evenly between fragments. The \polimi result corresponds to $x\sim2$, giving the light fragment twice as much energy as the heavy fragment which is not physically realistic. This discrepancy is a side effect of how \polimi samples each quantity independently and does not capture effects related to the de-excitation process. 

\figfirst{fig:W}(b) shows \Wtheta from each simulation divided by that for experiment. This ratio shows that, compared to the measured results, \polimi overpredicts \Wtheta by up to 90\% at low angles and underpredicts at high angles, while \cgmf and \freya underpredict at low angles and overpredict at high angles by a much smaller amount, about 10\% in each case. This discrepancy may indicate that \cgmf and \freya predict too many two-neutron events in which one neutron comes from one fragment and the other neutron from the complementary fragment, as opposed to both neutrons coming from the same fragment. 

This variation can be explored further by capturing the magnitude of the anisotropy as a one-dimensional parameter \Asym:
\begin{equation}
\label{eq:Asym}
\Asym = \frac{W(180)}{W(90)}\approx\frac{\Wth{175}}{\Wth{85}}
\end{equation}
Due to the \degrees{10} wide discretization of angles, the data are compared at \degrees{175} and \degrees{85}, which include pairs at $(\degrees{170},\degrees{180}]$ and $(\degrees{80},\degrees{90}]$, respectively.

\begin{figure}[!t]
	\centering
	\subfloat{\includegraphics[trim={0cm 0cm 0cm 0cm},clip,width=3in]{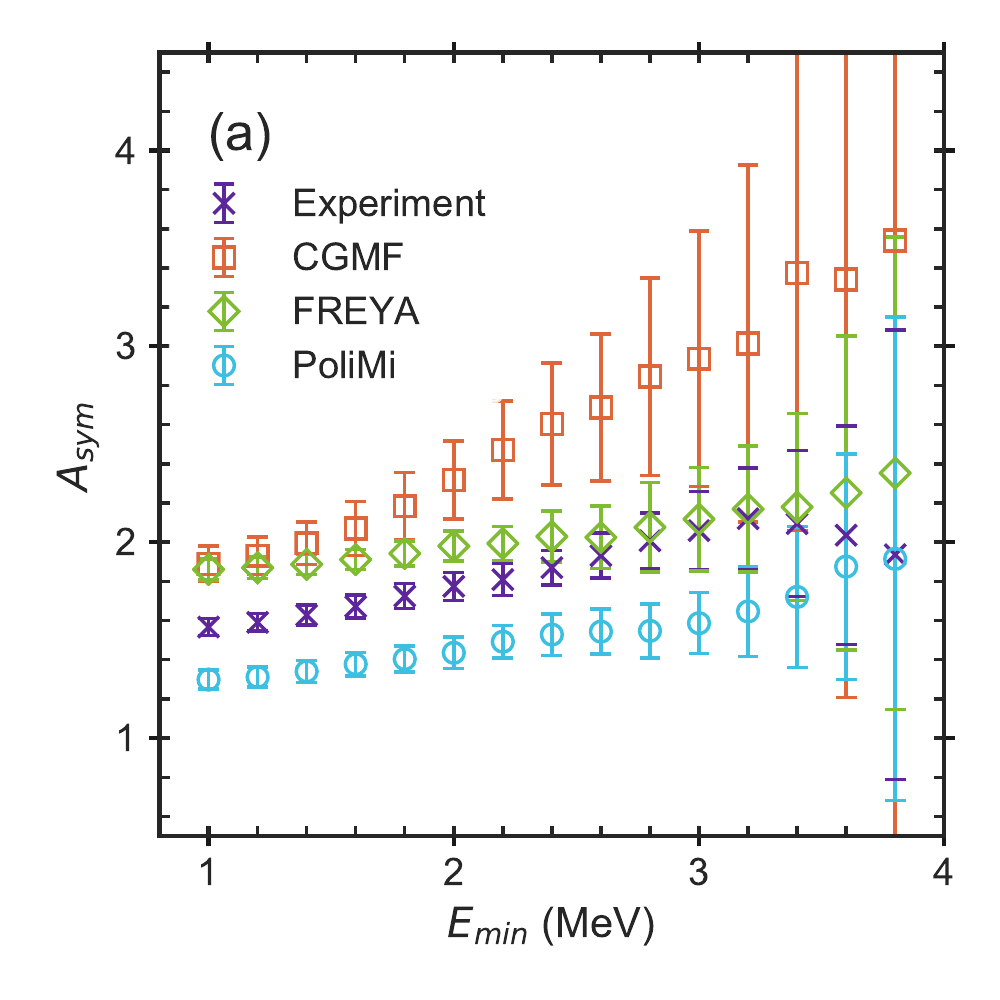}}
	\hfil
	\subfloat{\includegraphics[trim={0cm 0cm 0cm 0cm},clip,width=3in]{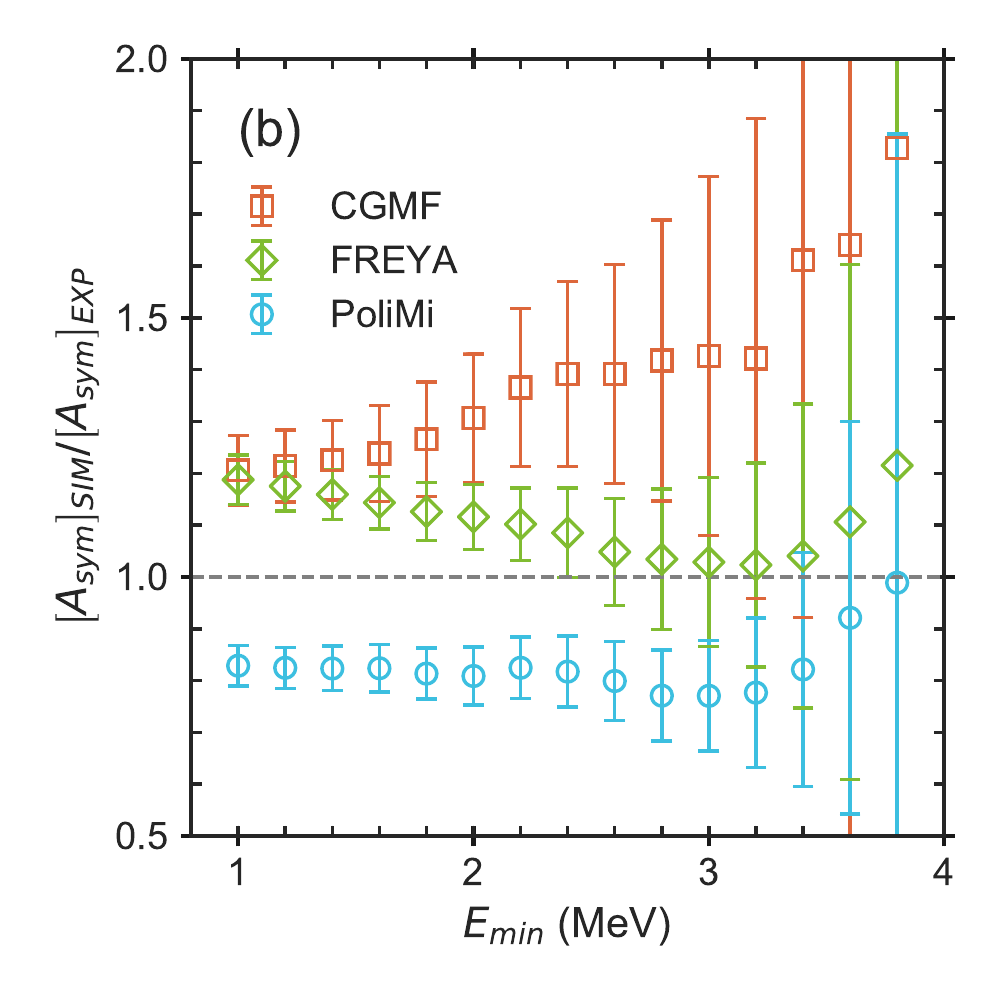}}
	\caption{\coloronline (a) Magnitude of the neutron emission anisotropy, \Asym, as a function of \Emin and (b) ratio between simulated results and measured data. The magnitude of anisotropy increases as the neutron population is limited to higher energies.}
	\label{fig:Asym_vs_Emin}
\end{figure}

The anisotropy in neutron energy can be observed by varying the neutron energy threshold, as shown in \fig{fig:Asym_vs_Emin}. The magnitude of the anisotropy increases as the energy threshold is increased and lower energy neutrons are omitted from the analysis. This increase occurs because neutrons detected at angles near \degrees{180} are likely emitted from different fission fragments in their direction of travel and therefore receive a boost in energy due to the frame of motion. This boost also occurs for neutron pairs emitted at \degrees{0}, however, the Chi-Nu array cannot identify events at \degrees{0} where two neutrons interact in the same detector. 
Neutrons detected at angles near \degrees{90} did not receive this boost, and therefore are emitted with a lower energy distribution. Thus, as the energy threshold is increased, events at angles near \degrees{90} are more likely to be removed from the population than events at \degrees{180}, thereby increasing \Asym.

\figfirst{fig:Asym_vs_Emin}(a) shows that \cgmf consistently produces the highest values of \Asym while \polimi consistently produces the lowest.  Note also that the uncertainties grow as \Emin increases because there are fewer events in the population, limiting statistics. \figfirst{fig:Asym_vs_Emin}(b) shows the ratio of each of the simulations to the experimental data. This ratio is roughly independent of $\Emin$ at $\sim0.8$ for \polimi, while \cgmf and \freya vary slightly as \Emin increases. The \freya ratio starts at $\sim1.2$ and drops toward to 1 as \Emin increases, while the \cgmf ratio starts at 1.2 and grows larger with increasing \Emin. % \rephrase{Although the \Asym values from experiment decrease above \MeV{3.2}, the changes are within the 1-$\sigma$ error bars, and therefore the behavior is not significant.}

\subsection{Neutron Energy Characteristics}

\begin{figure}[!t]	
	\centering
	\subfloat{\includegraphics[trim={0cm 0cm 0cm 0cm},clip,width=3in]{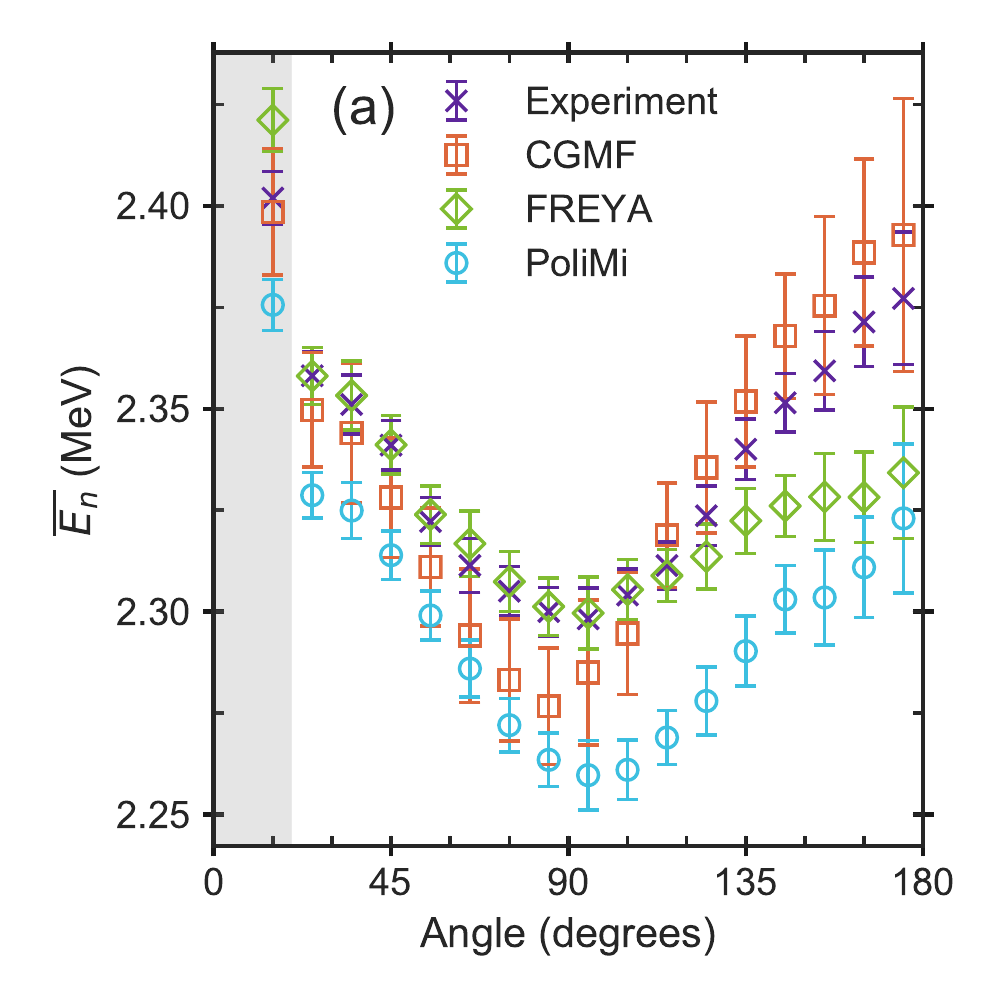}}
	\hfil
	\subfloat{\includegraphics[trim={0cm 0cm 0cm 0cm},clip,width=3in]{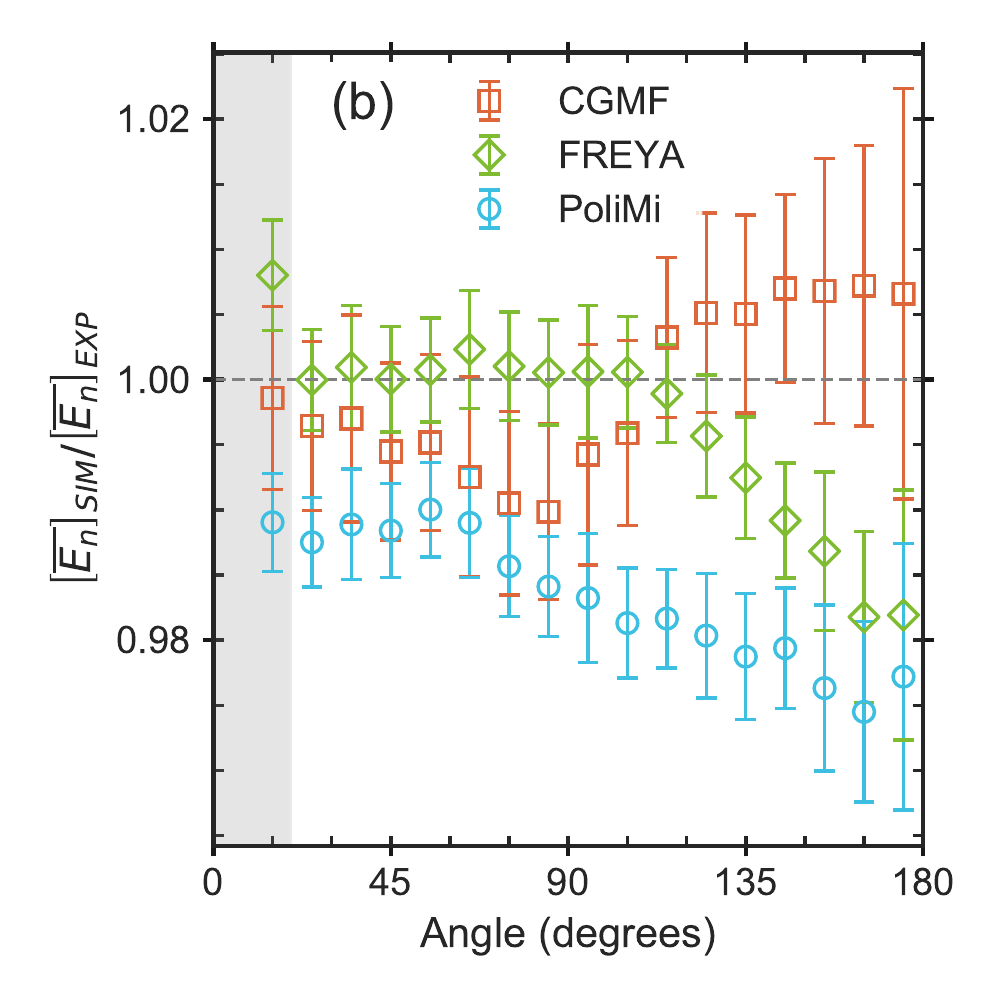}}
	\caption{\coloronline (a) Average neutron energy as a function of bicorrelation angle across a range of \MeV{1} to \MeV{4}. (b) Ratio between average simulated energy and average measured energy, demonstrating agreement within 3\% across all data. The gray region from \degrees{0} to \degrees{20} serves as a reminder that cross talk is significant over this range.}
	\label{fig:Esum}
\end{figure}

As stated in \secref{sec:asym}, the energies of prompt fission neutrons vary with their direction of emission relative to the direction of fission fragment motion. In detected bicorrelation events, this boost increases the average detected energies of pairs near \degrees{180}, which are likely to be emitted from opposite fragments in the fragment direction of motion. One can observe this effect by calculating the average neutron energy for all neutrons detected in bicorrelation events as a function of bicorrelation angle, defined as

\begin{equation}
\label{eq:Eave}
\Eave = \overline{(E_1+E_2)}/2.
\end{equation}
and shown in \fig{fig:Esum}(a). This distribution shows that, in all cases, the average neutron energy reaches a minimum near \degrees{90} and increases steadily until it reaches a local maximum at \degrees{180}. Note that \Eave is higher than expected at \degrees{15} due to cross-talk effects; the gray band for angles less than \degrees{20} serves as a reminder of this. 

Although the shapes are approximately the same in all cases for angles less than \degrees{90}, the behavior varies greatly above \degrees{90}. First, the minimum \Eave for \cgmf occurs at \degrees{85} while for all others it is at \degrees{95}. Second, \cgmf has the steepest increase in \Eave at angles up to \degrees{180}. Third, the experimental results are in excellent agreement with \freya at angles below \degrees{125}, but then the \freya \Eave levels out at higher angles while the \Eave of the data continues to rise. %\note{Ramona: Why does \freya level off?} 

\figfirst{fig:Esum}(b) shows the ratio between each simulation and experiment, demonstrating that the agreement among all results is very good, as all simulations are within 3\% of the experimental data. \polimi produces consistently lower energies than experiment while \cgmf produces lower energies below \degrees{135} and higher energies above \degrees{135}. \freya agrees with experiment below \degrees{125}, but it produces lower average energies above this angle.

\begin{figure}[!t]
	\centering	
	\subfloat{\includegraphics[trim={0cm 0cm 0cm 0cm},clip,width=3in]{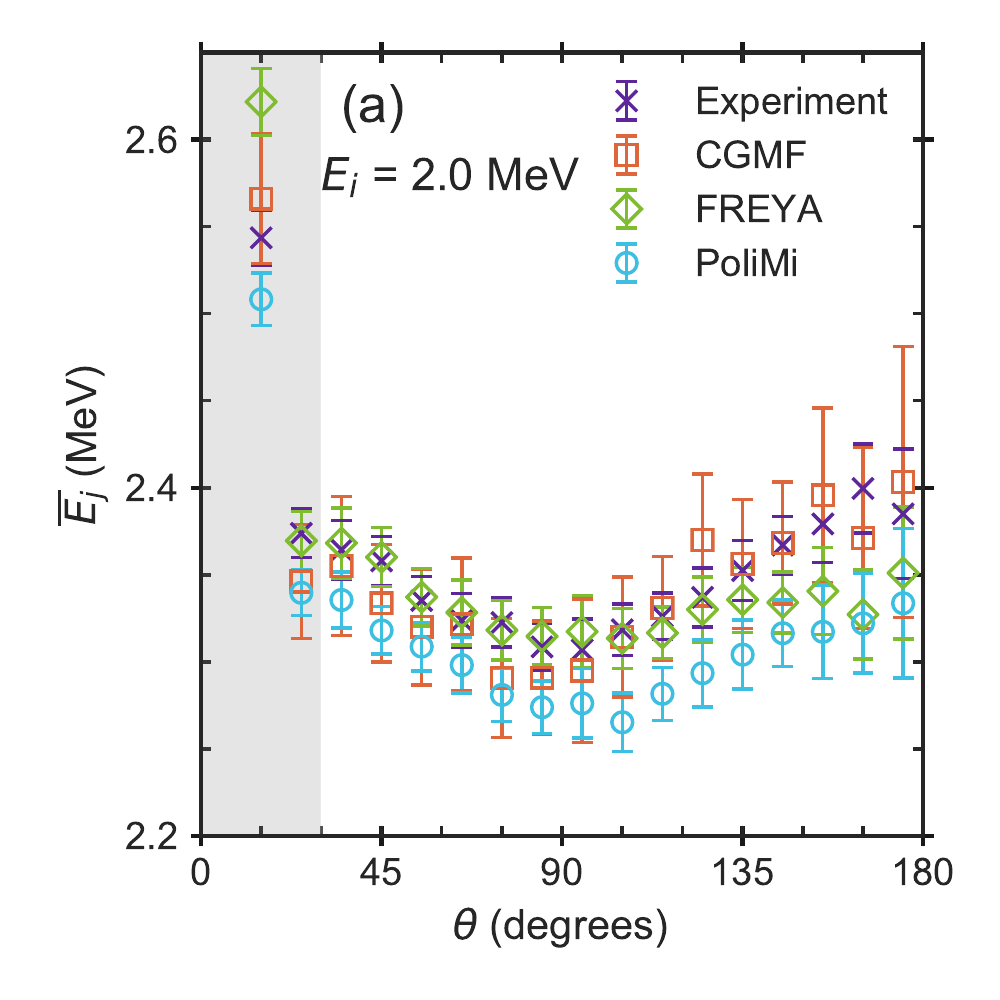}}
	\hfil
	\subfloat{\includegraphics[trim={0cm 0cm 0cm 0cm},clip,width=3in]{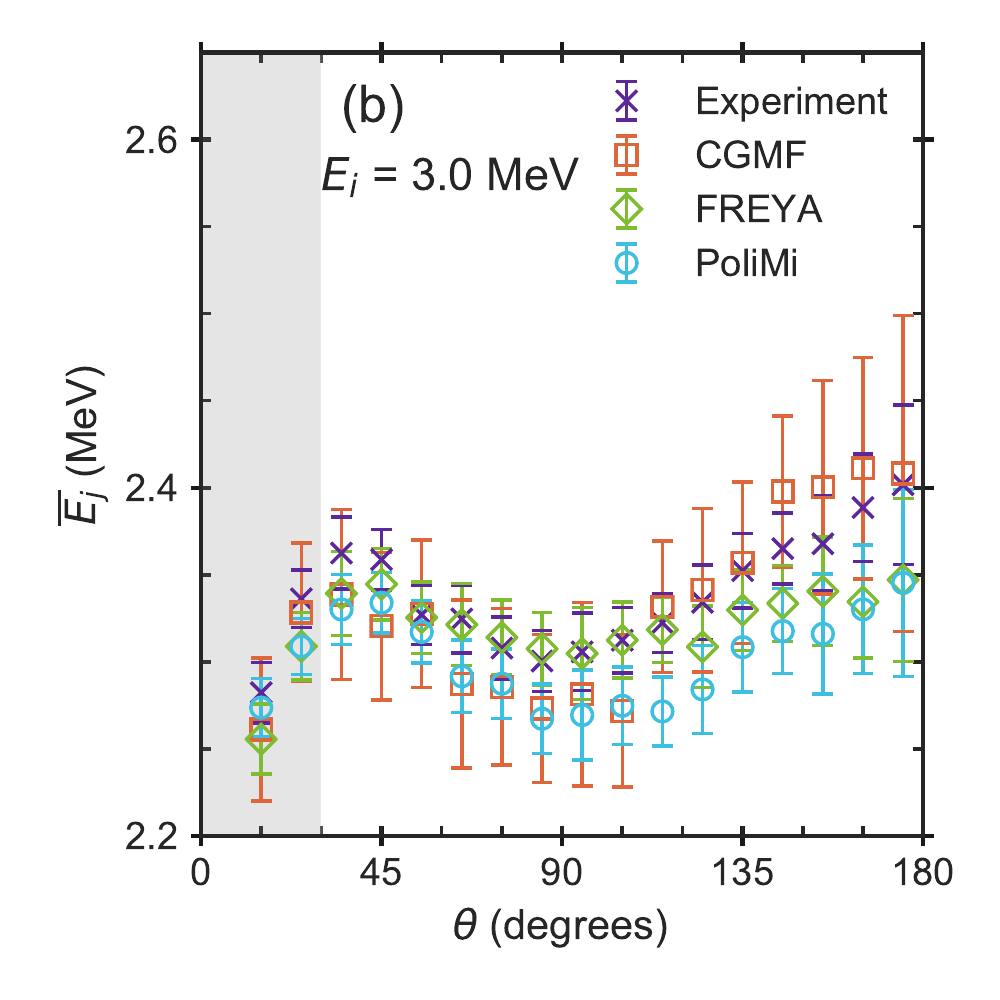}}
	\caption{\coloronline Average energy of neutrons detected in coincidence with a (a) \MeV{2} and (b) \MeV{3} neutron.}
	\label{fig:Ejave_vs_theta}
\end{figure}

While \fig{fig:Esum} provides a measurement of the energy distribution across the entire neutron population, it does not demonstrate whether the energy of one neutron depends on the energy of its bicorrelation partner. To determine this dependence, for fixed \Ei of \MeV{2} and \MeV{3}, the average energy \Ejave of the partner neutron is shown as a function of $\theta$ in \fig{fig:Ejave_vs_theta}. 

The distributions shown in \fig{fig:Ejave_vs_theta} share the same features as in \fig{fig:Esum}(a), although the behavior of data in angular bins less than \degrees{30} varies due to the dependence of cross talk on \Ei. In fact, no significant angular dependence was observed in the shape of \Eave or \Ejave at any \Ei. 

\begin{figure}[!t]
	\centering	
	\subfloat{\includegraphics[trim={0cm 0cm 0cm 0cm},clip,width=3in]{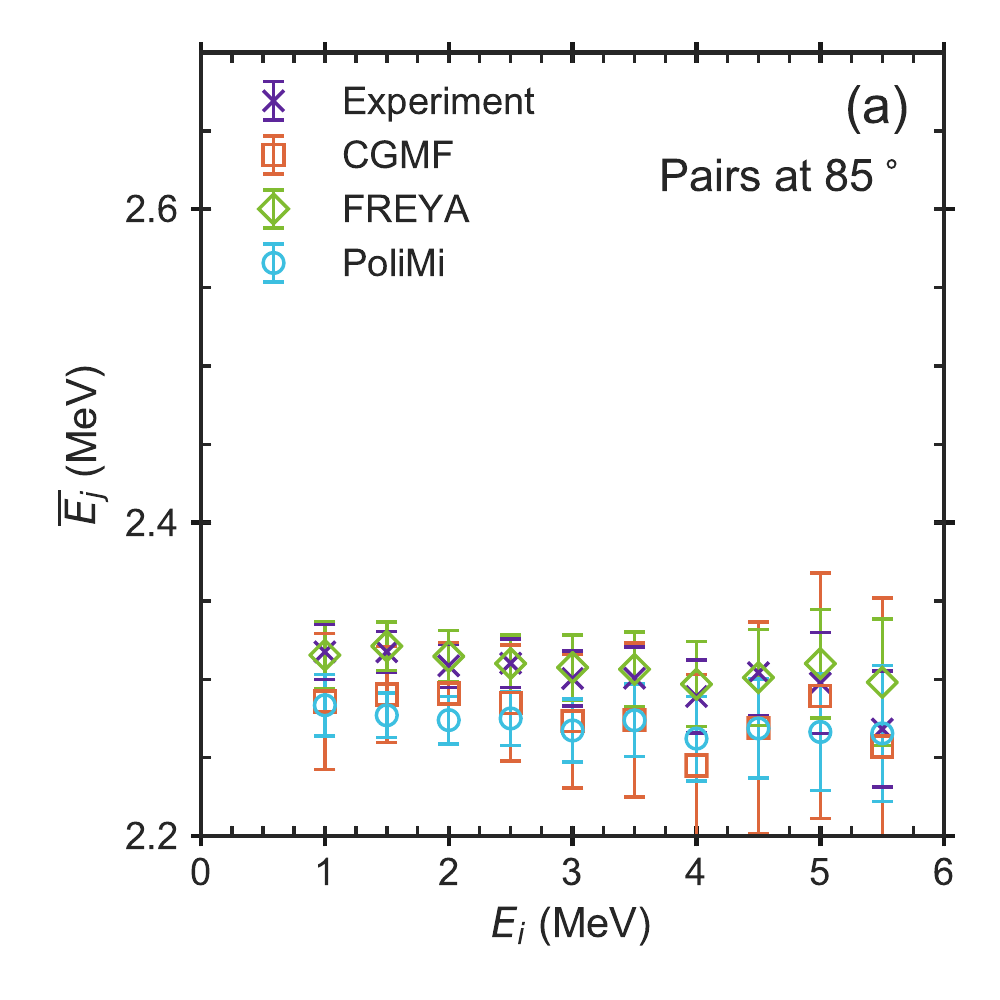}}
	\hfil
	\subfloat{\includegraphics[trim={0cm 0cm 0cm 0cm},clip,width=3in]{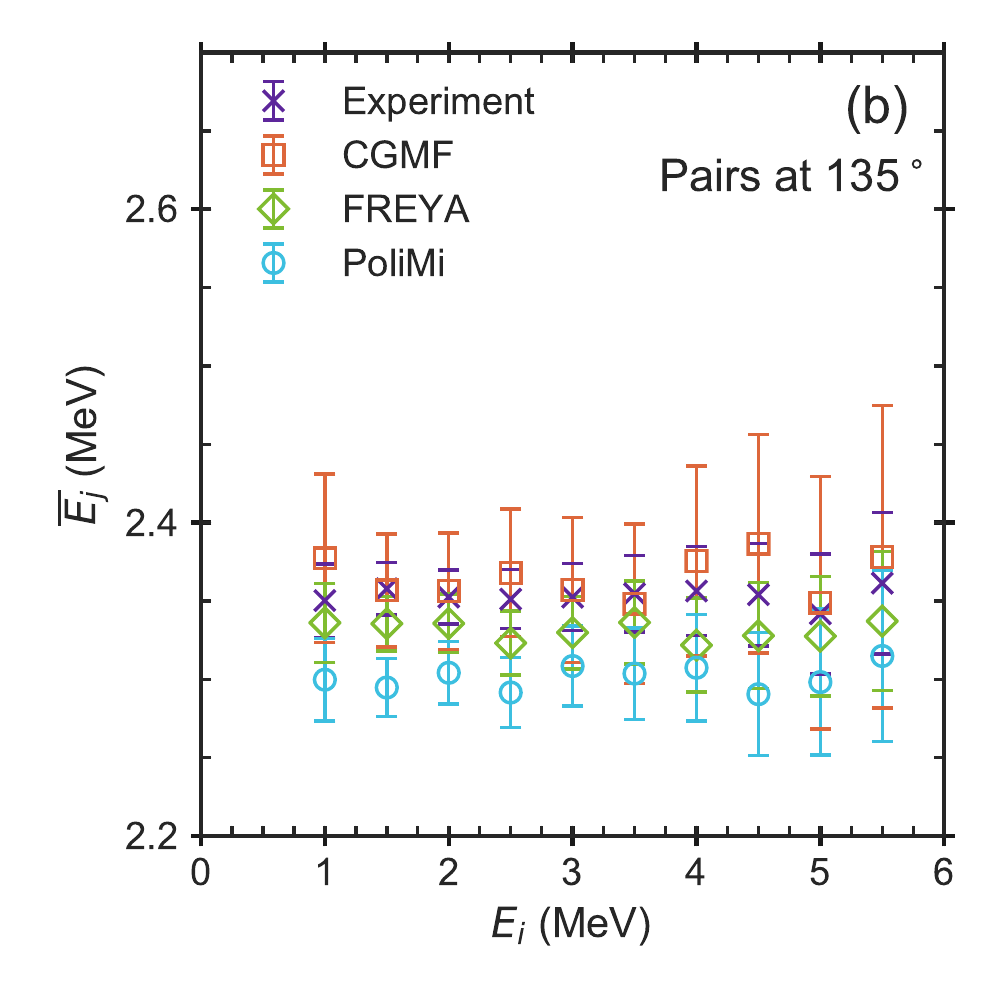}}
	\hfil
	\subfloat{\includegraphics[trim={0cm 0cm 0cm 0cm},clip,width=3in]{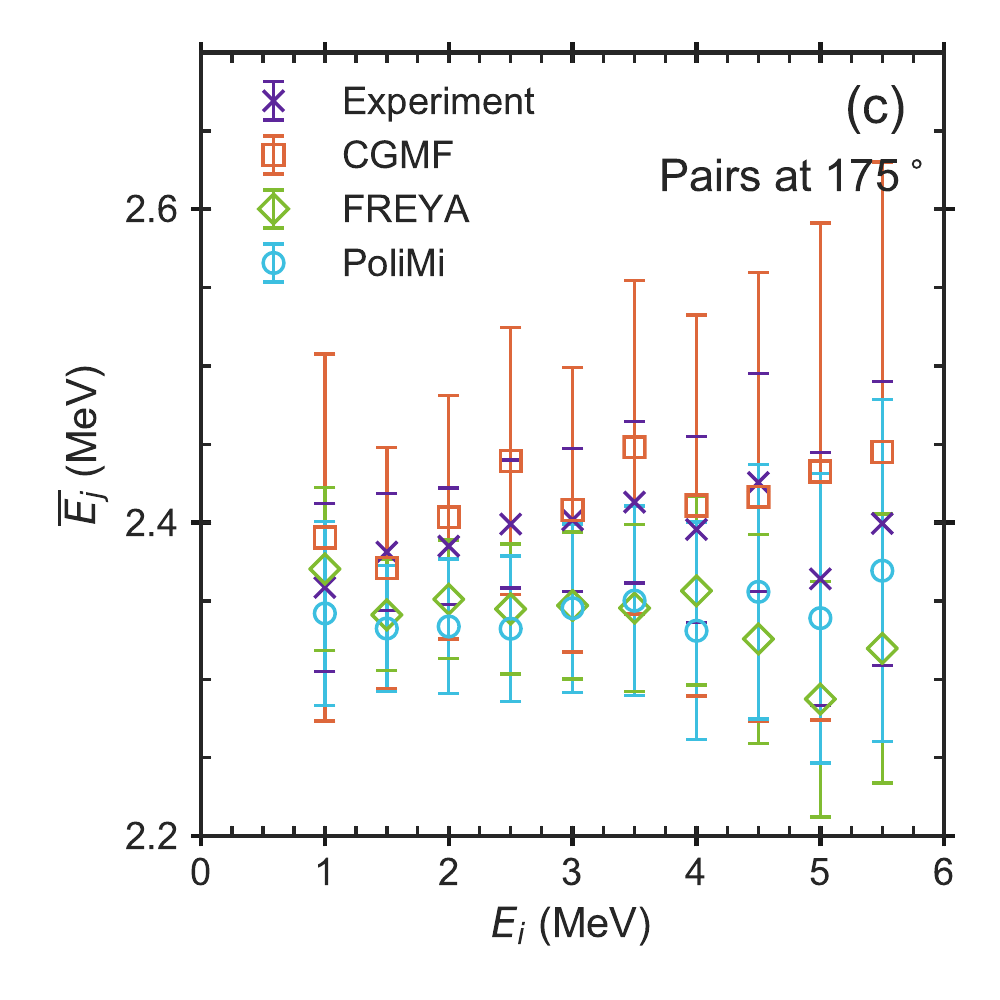}}
	\hfil
	\subfloat{\includegraphics[trim={0cm 0cm 0cm 0cm},clip,width=3in]{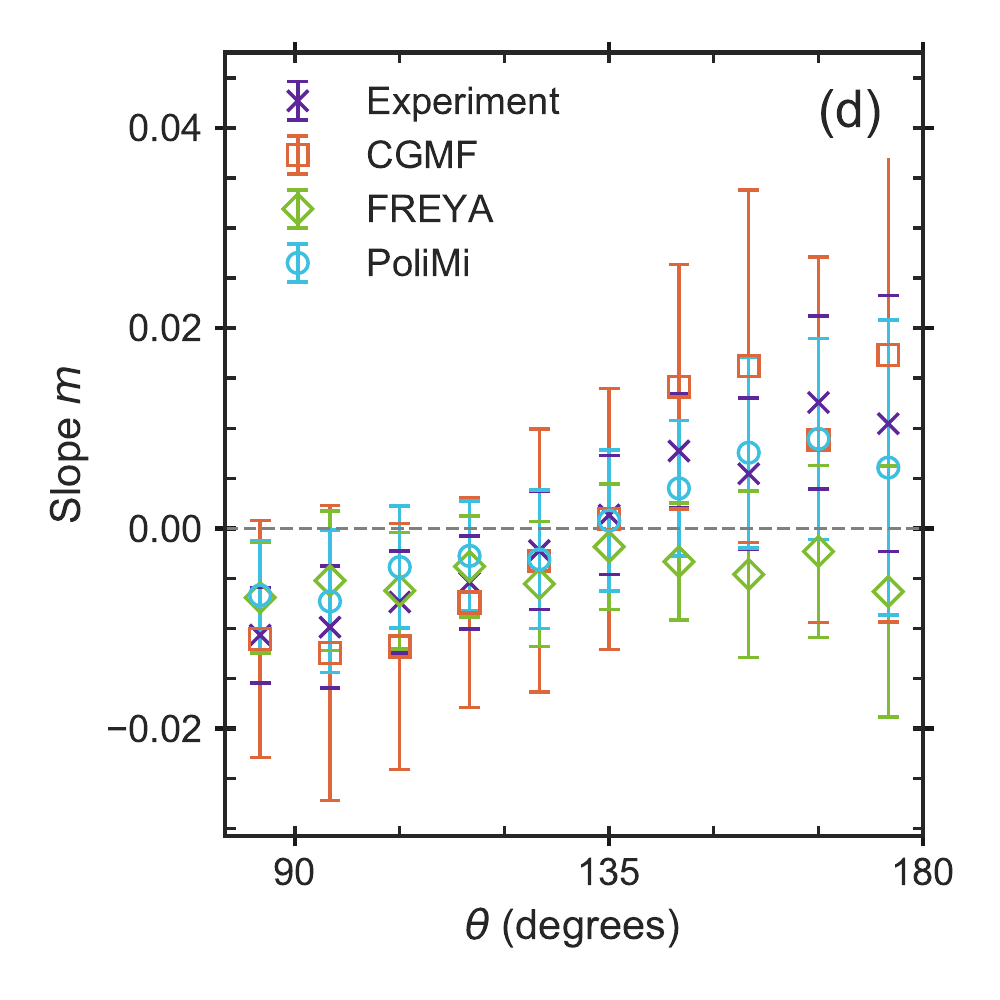}}
	\caption{\coloronline Average correlated neutron energy \Ejave for fixed energy \Ei for detector pairs at (a) \degrees{85}, (b) \degrees{135}, and (c) \degrees{175}, and (d) slope of least-squares fit to $\Ejave(\Ei)$ at angles \degrees{85} and higher.} 
	\label{fig:Ejave_vs_Ei}
\end{figure}

Some differences were seen, however, in the values of \Ejave as \Ei is varied. The dependence of \Ejave on \Ei can be enhanced by studying \Ejave as a function of \Ei at a fixed bicorrelation angle, as shown in \fig{fig:Ejave_vs_Ei}. 
% Higher \Ei interactions are more likely to produce a cross-talk event with a lower \Ej interaction, thereby lowering the overall \Ejave. 
\figfirst{fig:Ejave_vs_Ei} (a)-(c) shows $\Ejave(\Ei)$ at bicorrelation angles \degrees{85}, \degrees{135}, and \degrees{175}. While it is not immediately clear to the naked eye whether a dependence of \Ejave on \Ei exists, one can perform a least-squares linear regression on the data and determine whether there is a statistically significant nonzero slope, $m$, as shown in \fig{fig:Ejave_vs_Ei}(d). Angles below \degrees{85} are omitted, because cross talk was shown to be significant enough to contaminate the calculation of the slope at lower angles. %This range was selected by comparing the slopes calculated for the \polimi dataset with and without cross-talk events, which can be removed by MPPost in post-processing. 

There are several interesting aspects of this distribution. First and foremost, all four results have slopes within $2\sigma$ of $m=0.0$ across all angles. Thus, there is no statistically significant slope in any of the datasets. However, trends do exist in the data which will be discussed here and will be the subject of future work in order to reduce uncertainties and determine whether the trends are significant. The experimental data, as well as the \cgmf and \polimi simulations show a negative slope near \degrees{90}. Above \degrees{140}, the slope becomes positive, crossing zero near \degrees{135}. Since events with bicorrelation angle near \degrees{90} are likely emitted from the same fragment, the negative slope at angles near \degrees{90} could indicate that neutrons emitted from the same fission fragment compete with one another for energy. Events with neutrons emitted near \degrees{180} are likely to come from different fission fragments, indicating that there may be some positive correlation in neutron energies emitted from different fission fragments. Meanwhile, the \freya simulation results in negative slope across all angles, indicating that correlated neutrons produced by \freya may compete with one another for energy regardless of whether or not they were produced by the same fission fragment.

%\subsection{Systematic uncertainties}
%Although we performed some corrections, there exist others that we did not perform. What systematic uncertainties. We choose to show detected neutron pairs instead of emitted neutrons, so we don't perform all the corrections. (Difference of motivation)
%
%This section may need to go in the results section.
%
%Discuss systematic uncertainties or physical constraints on our system. 
%
%\begin{itemize}
%	\item Physical size of the detector $\rightarrow$ timing resolution, angular resolution
%	\item Distance to the detection $\rightarrow$ time of flight $\rightarrow$ timing resolution
%\end{itemize}
%
%Other things that are not included in simulation but are in measurement: Misclassification, overlap of fissions, threshold effects, neutrons being attributed to wrong fission

% \input{discussion}
\section{Conclusion}

This work investigated correlations in angle and energy between prompt neutrons emitted in the same \Cftft spontaneous fission event, including measuring the energy dependence between correlated neutrons for the first time. Experiments were performed using 42 components of the Chi-Nu detector array in a hemispherical configuration surrounding a fission chamber. %Measured events in which two or more prompt neutrons were detected in coincidence with a fission-chamber trigger were identified as ``bicorrelation'' events. 
The detector array was simulated in \mcnpxpolimi with three different fission models: \polimiipol, \cgmf, and \freya.
%the correlated neutron behavior in each model was compared to the data.

Characteristics of the correlated neutrons were studied with respect to the angle between the two neutrons. The large number of detectors produced a broad distribution of bicorrelation angles gathered into \degrees{10} bins. The \meters{1} flight path allowed for experimental timing resolution as low as \ns{1}, allowing excellent energy resolution to be attained for neutron energies between \MeV{1} and \MeV{4} from the time-of-flight calculations.

% NExt sentence is good, but repetitive with next paragraph, include 3% figure here to quantify it
The simulations showed good agreement with experiment for all measured quantities, while revealing interesting differences between fission event generators. The neutron emission anisotropy generated by \cgmf and \freya agreed within 10\% of experiment, while underpredicting the anisotropy at small angles and overpredicting it at high angles. On the other hand, \polimi showed poor agreement, differing up to 40\% from experiment at low angles. All simulated average neutron energies fell within 3\% of the experimental data. \freya produced the best agreement with experiment: the average neutron energies agreed with the data to 0.5\% for angles below \degrees{135}.

The average neutron energy was found to be negatively correlated with the energy of its correlated partner for pairs at \degrees{85}, indicating that neutrons may compete for emission energy at low angles, where neutrons are likely to be emitted from the same fission fragment. This correlation was found to be positive for pairs at \degrees{175}, where neutrons are likely to be emitted from different fission fragments. However, this result is inconclusive, because the uncertainties in the measurements result in  calculated slopes within $2\sigma$ of 0. Further experiments should be performed to study this effect in greater detail.

These conclusions lead to further questions that could be pursued by more sophisticated experiments. The ability to distinguish events with neutrons from the same fission fragment would determine of whether there is a competition for energy within the energy spectrum of the fragment, such as a reduction in the average emission energy for each subsequent neutron emission. Tracking the fission fragments would allow this analysis to be repeated with respect to the fission fragment motion. Extracting information about the neutrons at the time of their emission from the fragments, as opposed to relying on the information gleaned from the neutrons arriving at the detectors, which may have undergone some rescattering, would enable more direct comparison to the complete fission event models. Finally, repeating this measurement with \Putf(sf), with a neutron multiplicity closer to 2 $(\sim2.15)$ than \Cftft(sf) $(\overline \nu \sim3.76)$, would reduce the number of fission events with multiple neutron pairs in the same event. Thus, in this case, detected bicorrelation events are more likely to come from events where exactly two neutrons are emitted: either one from each fragment or two from the same fragment.

\section{Acknowledgments}

This work was performed under the auspices of the U.S. Department of Energy by Los Alamos National Security, LLC, under Contract No.~DE-AC52-06NA25396, by Lawrence Livermore National Security, LLC under Contract No.~DE-AC52-07NA27344, by Lawrence Berkeley National Laboratory under Contract DE-AC02-05CH11231, by the Consortium for Verification Technology under Contract No.~DE-NA0002534, by the Office of Nuclear Physics in the Office of Science under Contract No.~DE-AC02-05CH11231, and by the University of Michigan.

%%%%%%%%%%%%%% START BACKUP INFORMATION / APPENDICES %%%%%%%%%%%%%%%%
\section{References}
\bibliographystyle{elsarticle-num}
% \bibliography{library_omit_url} % Generated by my Mendeley account

\end{document}